\documentclass[useAMS,usenatbib,usegraphicx]{mn2e}
\usepackage[english]{babel}
\usepackage[utf8]{inputenc}
\usepackage{amsmath}
\usepackage{graphicx}
\usepackage{ulem}
\usepackage[colorinlistoftodos]{todonotes}
\usepackage{footnote}
\usepackage{tablefootnote}
\usepackage{supertabular}
\usepackage{times}
\usepackage{amssymb}
\usepackage{longtable}
\usepackage[varg]{txfonts}
\usepackage{graphicx}
\usepackage{natbib}
\usepackage{cancel}
\usepackage{soul}

\bibpunct{(}{)}{;}{a}{}{,}

\newcommand{\kms}{$\rm km\,s^{-1}$}

\newcommand{\hi}{\mbox{H\,\sc{i}}}
\newcommand{\hii}{\mbox{H\,\sc{ii}}}

\newcommand{\g}{G25.8700+0.1350}

\title[]{A multi-wavelength analysis of the diffuse \hii\, region G25.8700+0.1350}
 
\author[S. Cichowolski et al.] {S. Cichowolski$^1$\thanks{Member of
    the Carrera del Investigador Cient\'{\i}fico of CONICET,
    Argentina.}, N. U. Duronea$^{2\,\star}$, L. A. Suad$^{2\,\star}$, E. M. Reynoso$^{1\,\star}$, R. Dorda$^{3}$\\
$^{1}$ Instituto de Astronom\'{\i}a y F\'{\i}sica del Espacio (UBA, CONICET), CC 67, Suc. 28, 1428 Buenos Aires, Argentina\\
$^{2}$ Instituto Argentino de Radioastronomía (CCT-La Plata, CONICET; CICPBA), C.C. No. 5, 1894,Villa Elisa, Argentina\\
$^{3}$ Departamento de F\'{\i}sica, Ingenier\'{\i}a de Sistemas y Teor\'{\i}a de la Se\~nal, Universidad de Alicante, Carretera de San Vicente del Raspeig, E03690 Alicante, Spain\\}

\begin{document}

\date{}

\pagerange{\pageref{firstpage}--\pageref{lastpage}} \pubyear{}

\maketitle

\begin{abstract}
We present a multiwavelength investigation of the \hii\, region \g, located in the inner part of the Galaxy. In radio continuum emission, the region is seen as a bright arc-shaped structure. An analysis of the H{\sc i} line suggests that \g\, lies at a distance of 6.5 kpc. The ionized gas is bordered by a photodissociation region which  is encircled by a molecular structure where four molecular clumps are detected. At infrared wavelengths, the region is also very conspicuous.
Given the high level of visual absorption in the region, the exciting stars should be searched for in the infrared band. In this context, we found in the literature one Wolf-Rayet and one red supergiant which, together with 37 2MASS sources candidates to be O-type stars,  could be related to the origin of \g.
Finally, as expanding \hii\, regions are hypothesized to trigger star formation, we used different infrared point source catalogues to search for young stellar object candidates (cYSOs). A total of 45 cYSOs were identified projected onto the molecular clouds.

\end{abstract}
\label{firstpage}


\section{Introduction}

Massive stars are known to play an important role in disrupting, modifying,  and dispersing the ambient molecular gas through their ultraviolet (UV) radiation, strong  winds, outflows, and eventually, with  supernova explosions. The strong UV radiation of massive stars, ionizes the molecular gas creating H{\sc ii} regions, which are expected to expand in the interstellar medium because of the high difference in pressure between the ionized and the ambient neutral gas. This can originate large distortions in their surroundings and even compression of nearby molecular clouds stimulating the formation of a new generation of stars \citep{elm77,lef94,zin07}.

The feedback of massive stars and the evolution of their associated  H{\sc ii} regions determine the physical conditions of their environs and the star formation rate in the region.  Hence, it is important to study the interstellar medium  adjacent to Galactic \hii\, regions  since they can provide substantial information, not only about the physical conditions where massive stars are born, but also to set empirical constraints to existing theoretical models and to improve our knowledge of the physical processes leading to stellar formation and evolution.

In this paper we present the very first multiwavelenght analysis of the diffuse \hii\, region catalogued by \citet{hel06} as \g. This region consists in a large, bright partial arc, located in the inner Galaxy, near the tangent point as inferred from observations in several radio recombination lines (RRL). This \hii\, region is especially interesting because it is close to two young high-mass clusters, RSGC1 \citep{fig06,dav08} and RSGC2 \citep{dav07}, which are dominated by red supergiants (RSGs), evolved high-mass stars. These clusters are separated from the center of \g\, by $39'8$ and $22'6$ respectively. Moreover, the largest concentration of red supergiants (RSGs) known in our Galaxy \citep{neg12,dor16} can be found around these clusters. According to the high density of RSGs, this region probably is one of the most intense star-forming places in the Galaxy.

\section{Data sets}

To carry out this study we made use of public archival data from near-IR to radio wavelengths. In what follows, we describe the datasets employed.

\begin{itemize}

\item Radio continuum data at 1420 MHz were extracted from the Multi-Array Galactic Plane Imaging Survey \citep[MAGPIS; ][]{hel06}. This survey was constructed combining Very Large Array (VLA) observations in the B, C, and D arrays. VLA data have a limited $u,v$ coverage, and are insensitive to structures much larger than 1 arcmin. Hence, the MAGPIS images include short spatial frequencies  obtained from observations with the 100-m Effelsberg radiotelescope so as to sample all diffuse, extended structures. The angular resolution is 6\hbox{$.\!\!{}^{\prime\prime}$}2 $\times$ 5\hbox{$.\!\!{}^{\prime\prime}$}4, and the sensitivity, 0.2 mJy beam$^{-1}$.\\

\item IR data were obtained from the Herschel infrared Galactic Plane Survey (Hi-Gal). Hi-Gal \citep{mol10} used the Photodetector Array Camera and Spectrometer \citep[PACS;][]{pog10} and the Spectral and Photometric Imaging Receiver \citep[SPIRE;][]{gri10} cameras. The instruments detecting emission at 70, 160 $\mu$m (PACS) and 250 $\mu$m (SPIRE) have angular resolutions of 5\hbox{$.\!\!{}^{\prime\prime}$}5, 12\arcsec, and 17\arcsec, respectively. We obtained the UNIMAP level 2.5 images using the Herschel Science Archive.\footnote{http://www.cosmos.esa.int/web/herschel/science-archive}.
We complemented IR emission using data from the Spitzer Space Telescope. We used the emission at 8 $\mu$m from the Infrared Array Camera \citep[IRAC;][]{wer04}, and 24 $\mu$m from the Multiband Imaging Photometer \citep[MIPS;][]{rie04}. The spatial resolutions are 2\arcsec and 6\arcsec\, for 8 and 24 $\mu$m, respectively.
\\

\item We used \hi\, 21 cm data from the VLA Galactic Plane Survey \citep[VGPS; ][]{VGPS}. As in the case of MAGPIS, the interferometric data must be completed with short-spacings to sample extended structures. In this survey, single dish data were supplied by the Green Bank Telescope (GBT).  The \hi\, line data of the VGPS have an angular resolution of $1^\prime \times 1^\prime$ and a spectral resolution of 
1.56 km s$^{-1}$, although channels are sampled each 0.824 km$^{-1}$ for consistency with the Canadian Galactic Plane Survey (CGPS).\\

\item To survey the molecular emission, we used $^{13}$CO(1-0) line data obtained from the Boston University-FCRAO Galactic Ring Survey  \citep[GRS\footnote{http://www.bu.edu/galacticring/new\textunderscore index.htm};][]{jac06} carried out with the SEQUOIA multi-pixel array receiver on the FCRAO 14 m telescope.  This survey covers the Galactic plane in the range $18\fdg0 < l < 55\fdg7$ and $-1\fdg0 < b < 1\fdg0$, and has a sensitivity $<$ 0.4 K. The angular resolution and sampling are 46$"$ and 22$"$, respectively. The velocity resolution is 0.2 \kms , covering a range from -5 \kms\ to +135 \kms, respectively. We also used CO(3-2) line data obtained from the CO High-Resolution Survey \citep[COHRS\footnote{http://dx.doi.org/10.11570/13.0002};][]{demp13}. The data were taken using the Heterodyne Array Receiver Programme on the James Clerk Maxwell Telescope (JCMT). The angular resolution and sampling are 14$"$ and 6$"$, respectively. The velocity resolution is 1 \kms, spanning from --30 \kms\ to +155 \kms, respectively.\\

\item
We used 1.1mm continuum data from the Bolocam Galactic Plane Survey (BGPS)\footnote{http://irsa.ipac.caltech.edu/Missions/bolocam.html}. This survey is contiguous over the range $-10\fdg$5 $\leq$ $l$ $\leq$ 90$\fdg$5, $|b|$ $\leq$ 0$\fdg$5, and is extended to $|b| \leq 1\fdg$5 for 75$\fdg$5 $\leq$ $l$ $\leq$ 87$\fdg$5. Additional cross-cuts were performed at $l$ = 3$\fdg$0, 15$\fdg$0, 30$\fdg$0, and 31$\fdg$0, and four supplementary regions of the outer Galaxy, covering 170 deg$^2$ in total \citep{agui11}. Bolocam is a facility consisting of a 144-element bolometer array camera mounted at the Cassegrain focus of the 10.4m mirror of the CSO on the summit of Mauna Kea. The array field of view (FOV) of the instrument is 7'.5, and its effective resolution is 33$^{\prime\prime}$ \citep{glen03}.

\end{itemize}

\section{\g\, and its local ISM}

\begin{figure}
\centering
\includegraphics[width=9.5cm]{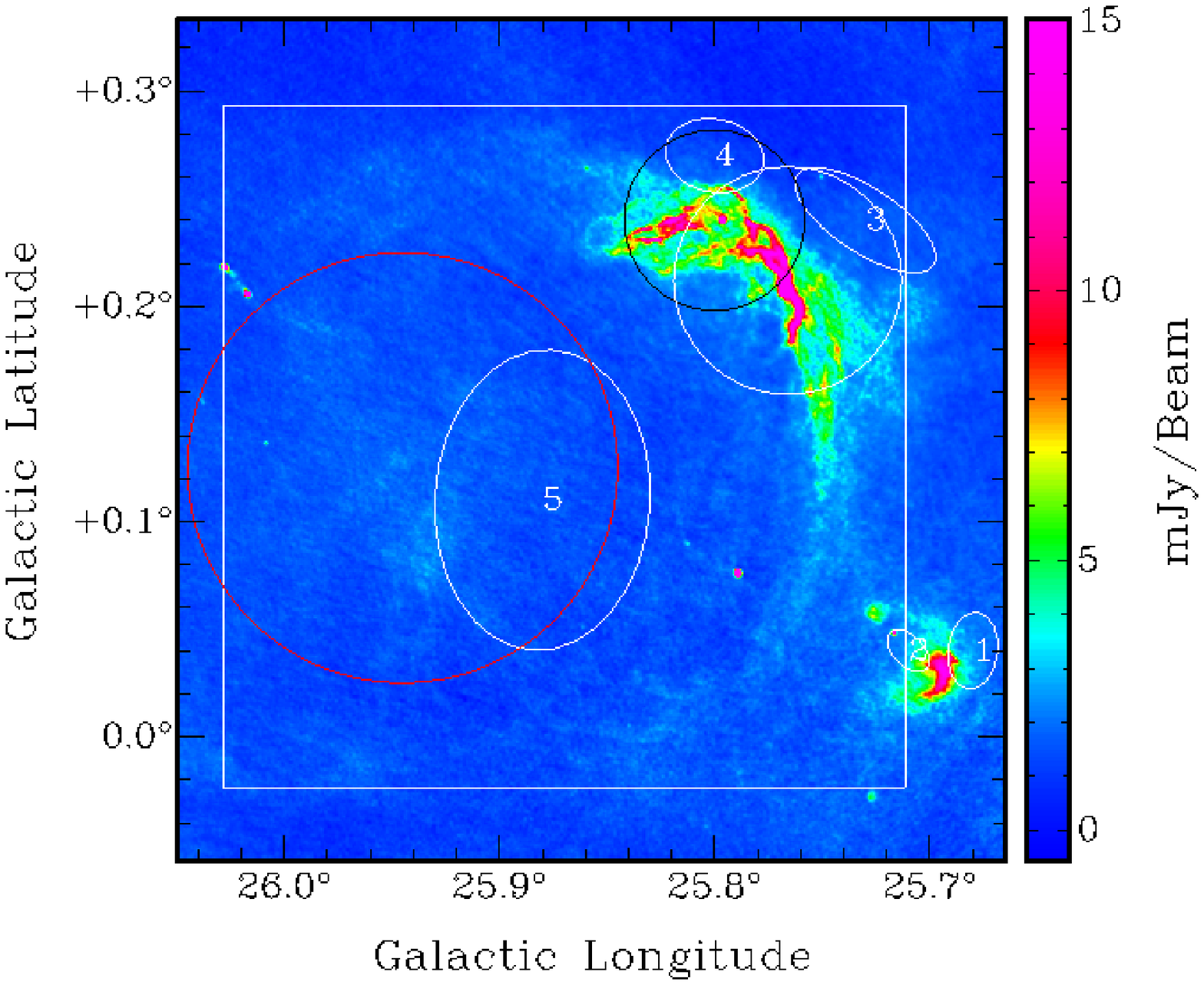}
\caption{Radio continuum image of the region at 1420 MHz obtained from MAGPIS. The  box indicates the location of \g. Red, black, and white circles indicate the locations where RRL were detected by  \citet{loc96}, \citet{sew04}, and \citet{quia06} respectively. The size of the circles correspond to the beam size of each observation. The ellipses, numbered as in Table \ref{andco}, enclose the ionized regions having CO associated, as indicated by \citet{anda09}.}
\label{continuo}
\end{figure}

Figure \ref{continuo} shows the emission distribution of the region under study at 1420 MHz.  \g\, is the
 large,  bright source, whose location and size, as given by \citet{hel06}, are indicated by the box.
Several  RRLs were detected in the region (see Fig. \ref{continuo}). \citet{loc96} detected the 6 cm (H109$\alpha$ and H111$\alpha$)  RRL  at ($l, b$) = (25\fdg945, +0\fdg125) at the velocity of $104.0 \pm 1.1$ \kms. The line 110 H$\alpha$ was detected at  ($l, b$) = (25\fdg8, +0\fdg24) at the velocity of 112.1 $\pm$ 0.1 \kms\, by \citet{sew04}.
Finally, \citet{quia06} detected towards ($l, b$) = (25\fdg766, +0\fdg212), the 91$\alpha$ and 92 $\alpha$ lines of the H, He and C  at 111.44 $\pm$ 0.17 \kms, $111.35 \pm 1.89$ \kms\, and 124.52 $\pm$ 0.96 \kms, respectively.

Since \g\, is located in the first Galactic quadrant, two distances, near (N) and far (F), are possible for each radial velocity up to the tangent point. Resolving the kinematic distance ambiguity (KDA)  is not easy, specially when the radial velocity of the source is very close to the velocity of the tangent point (T). The most straightforward method to solve the KDA for \hii\, regions consists in constructing a  21 cm \hi\, absorption spectrum toward the radio continuum emission and compare the absorption features with corresponding \hi\, emission peaks, where the last emission feature detected indicates the velocity of the tangent point. The detection of \hi\, absorption up to the tangent point implies that the source lies at the far distance inferred from the RRL velocity. Otherwise, it is safe to assume that the \hii\, region is located at the near distance.

In the direction of \g\,, the velocity of the tangent point is about +125 \kms , and the corresponding kinematic distance, around 7.7 kpc. Several attempts are found in the literature to solve for the KDA in \g\,. Applying the canonical method described above, \citet{quib06} inferred that the region is at 7.3 kpc, which corresponds to the near distance. On the other hand, \citet{sew04}  used  H$_2$CO line observations to disentangle the KDA problem for several HII regions; for \g, they cannot unambiguously distinguish between the near and far position. Finally, \citet{anda09} resolved the KDA using \hi\, and $^{13}$CO surveys for 266 HII regions located in the inner Galaxy with RRL emission detected and with molecular gas associated \citep{andb09}. In addition to performing the usual absorption analysis, they searched for self-absorption by comparing the \hi\, and $^{13}$CO lines, where the signature of self-absorption is the presence of an \hi\, absorption feature together with molecular emission at the same velocity.  

In Table \ref{andco}, we summarize the relevant results reported by \citet{anda09}. In the area of \g\, the authors identified five molecular clouds related to ionized gas,  which are indicated by grey ellipses in Fig. \ref{continuo} and described in Table \ref{andco}. Column 2 gives the ID of the \hii\, regions related to each cloud as given by \citet{andb09}, while their Galactic coordinates are given in columns 3 and 4. Column 5 gives the radial velocity of the detected RRL. Columns 6 and 8 show the result (N, F, T) obtained according to each method, either  the conventional \hi\, absorption analysis (dubbed 'EA' for Emission/Absorption) or the search for self-absorption ('SA').  Columns 7 and 9 give their corresponding quality parameters, where 'A' and 'B' stand, respectively, for high and low confidence of the results. Finally, the suggested distance is given in column 10. As can be seen in Fig. \ref{continuo}, clouds 1 and 2 are related to another \hii\, region \citep[cataloged as G\,25.7056+0.0389 by][]{hel06}, while clouds 3, 4, and 5 are probably associated with \g\, since they share both the radial velocity and spatial location.
It calls the attention, however, that while regions 3 and 4 are put at the near distance, region 5 seems to be, although with less reliability, at the far side.

In summary, even though the distance to \g\, has been estimated by many authors,  there is no agreement in the different results they achieve. Since having an accurate distance is crucial for determining the physical properties of the region, we will obtain our own estimation based on the radio continuum and \hi\, available data.

\begin{table*}
\caption{Parameters of \hii\, regions associated with CO clouds. } 
\label{andco}
\centering  
\begin{tabular}{l c c c c c c c c c }
\hline \hline
&ID & $l$ (deg.) & $b$ (deg.) & V$_{\rm LSR}$ (\kms)  & EA & Q-EA & SA & Q-SA& D (kpc) \\
\hline 
1&U25.72+0.05a&25.72 & 0.05& 53.3& --& --& N& B & 3.6\\
2&U25.72+0.05b&25.72 & 0.05&102.0& --&--& F& A& 9.2\\
3&C25.77+0.21& 25.77&0.21& 110.8&T& --& N& A& 6.7\\
4&U25.80+0.24& 25.80& 0.24& 112.1 & T&--&N&A& 6.8$^{\rm a}$ \\
5&D25.94+0.10&25.94&0.10& 104.0& T&--& F&B&9.1\\
\hline
\end{tabular}
\begin{list}{}{}
\item{$^{\rm a}$}
 The distance information given for this source is confusing since its radial velocity was not properly taken from the work of \citet{andb09}. We adopt the near distance associated with  V$_{LSR}$ = 112.1 \kms\, instead of the one given by \citet{anda09}, i.e. 7.7 kpc.
\end{list}
\end{table*}

\subsection{Distance estimation}\label{dist}

\begin{figure}
\centering
\includegraphics[width=9cm]{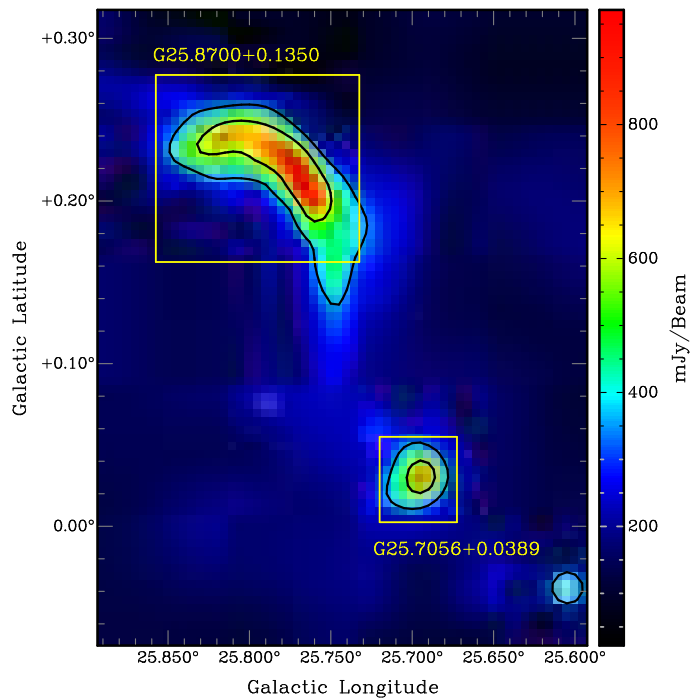}
\caption{MAGPIS image of the region at 1420 MHz convolved to a 60$^\prime \times 60^\prime$ beam in order to match the resolution of the \hi \ data. The intensity scale is displayed by a color bar to the right. The two yellow boxes indicate the regions used to compute the absorption profiles towards \g\, and G25.7056+00389, shown in Fig. \ref{abs}. Black contours at 350 and 600 mJy beam$^{-1}$ are plotted.}
\label{boxabs}
\end{figure}

\begin{figure}
\centering
\includegraphics[width=8.5cm]{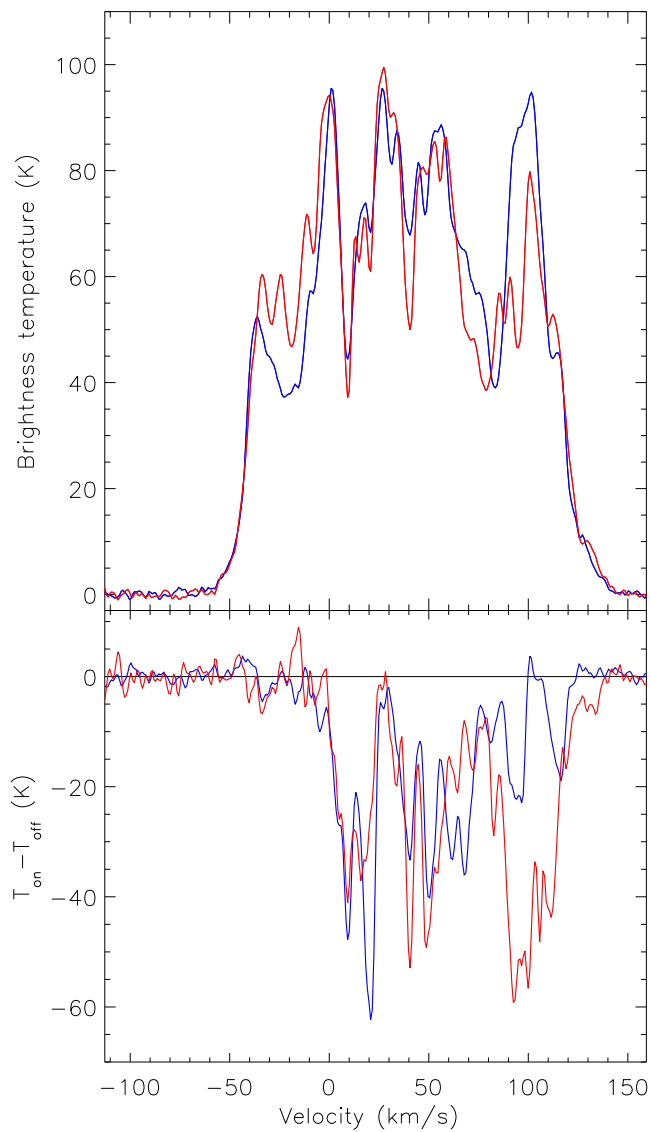}
\caption{\hi \ emission-absorption pairs towards \g \ (blue lines) and G25.7056+00389 (red lines). Off-source \hi \ profiles were computed employing a bilinear fit as explained in the text.}
\label{abs}
\end{figure}

In what follows, we perform an \hi \ absorption study of \g \ with the aim of constraining its systemic velocity and distance. 
For consistency, we have convolved the
1420 MHz continuum image obtained from MAGPIS down to the resolution of
the \hi \ data.
The on-source (T$_{\rm {on}}$) absorption profile was obtained towards the brightest region of \g,
enclosed by the radio continuum isophote at 600 mJy beam$^{-1}$. To obtain the
expected \hi \ profile at this region (T$_{\rm {off}}$), we performed a bilinear fit using only the
\hi \ pixels within the yellow box in Fig. \ref{boxabs} outside the outer radio
continuum contour, at 350 mJy beam$^{-1}$. The method \citep[e.g.][]{rcw17} consists in replacing
those pixels within the inner contour by the average between two linear fits: one
 along Galactic Latitude and the other along Galactic Longitude. The result is
shown by the blue solid line in Fig. \ref{abs}. The absorption feature at the
highest positive value appears at $\sim +115$ km s$^{-1}$. No absorption features
are seen at negative values.

To facilitate the analysis of the \hi \ absorption profile associated with \g, we
followed the same method described above and obtained a second profile towards
the bright, nearby \hii \ region G25.7056+00389. The result is displayed by
the red solid lines in Fig. \ref{abs}. A comparison between both profiles readily
shows that although the \hi \ emission level at $\sim +125$ km s$^{-1}$ is the same for
both sources, the red absorption profile
contains a $\sim 10$ km s$^{-1}$-width feature which is absent in the blue
profile. We thus interpret that the last absorption feature towards \g \ occurs before reaching the tangent point, and hence the region lies in the near side of the two possible distances associated with $v=+115$ km s$^{-1}$: 6.5 and 8.7 kpc, according to the Galactic rotation model of \citet{fbs89}. In what follows, we will adopt a distance of 6.5 $\pm$ 1 kpc for \g.

\subsection{1420 MHz emission}

\subsubsection{Radio continuum}\label{seccont}

Figure \ref{continuo} clearly shows that \g\, presents a very bright arc-shape structure superposed onto a faint, more extended emission. We estimated the average flux density of this emission to be 1.3 mJy beam$^{-1}$.  Accounting for this background, we obtained that the total flux density of the bright arc is $S_{\rm 1420} = 12 \pm 1$ Jy, where the uncertainty involves any arbitrary assumption in determining the limits of the region.

We then estimate the physical parameters of \g\, using the model developed by \citet{mez67}.
We consider the spherical source model, with an angular diameter of $\theta_{\rm sph} = 20\arcmin$ and a filling factor of 0.08. 
Assuming our estimated distance value D = $ 6.5 \pm 1.0$ kpc, and the electron temperature inferred by \citet{quib06}, $T_{\rm e} = 6120 \pm 100$ K, we obtain that the region contains $2500 \pm 1110$ M$_{\odot}$ of hydrogen ionized mass and an electron density of n$_{\rm e} = 43 \pm 10$ cm$^{-3}$, which is similar to the value n$_{\rm e} = 45.7 $ cm$^{-3}$ obtained by \citet{quib06} using RRL data.

Finally, we can estimate the number of UV photons necessary to keep the gas ionized, using the relation

$ N_{\rm UV} = 0.76 \times 10^{47} \, (\frac{Te}{10^4 K})^{-0.45}\, (\frac{\nu}{\rm GHz})^{0.1}\, (\frac{D}{kpc})^2\, (\frac{S_{\nu}}{Jy})$ s$^{-1}$  \citep{cha76}, which yields $ N_{\rm UV} = (5.0 \pm 1.6) \times 10^{49}$ s$^{-1}$ for \g.

\subsubsection{HI line emission distribution}\label{secthi}

\begin{figure}
\centering
\includegraphics[width=9cm]{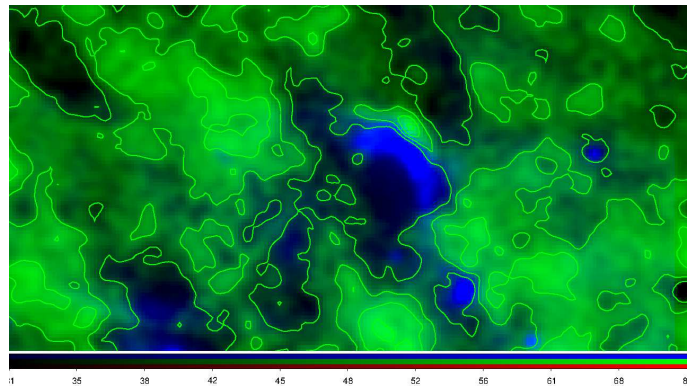}
\caption{Green: \hi\, emission distribution averaged in the velocity range from 109.8 to 118.1 \kms. Contour levels are at 40, 50, 60, and 70 K. Blue: radio continuum emission at 1420 MHz.}
\label{hi}
\end{figure}

We have inspected the \hi\, emission distribution in the velocity range where the RRLs were observed and  noticed two striking features in the velocity interval from 109.8 to 118.1 \kms\ (Fig. \ref{hi}): a)  a conspicuous \hi\, emission  minimum  around ($l$, $b$)= (25\fdg8, 0\fdg2), and b) a bright \hi\, feature observed next to the arc-shaped structure detected in the radio continuum emission at 1420 MHz (see Sect. 3.2.1).  Both features strongly suggest that the massive stars have ionized and probably swept up the surrounding ionized and neutral gas, generating an irregular cavity \citep{wea77}. The morphology of \g\ could be explained then as an \hii\ region bounded by density toward higher Galactic longitudes, and bounded by ionization toward lower Galactic longitudes, probably as a result of the action of the powering stars over an in-homogeneously  distributed original  neutral gas. This scenario is in agreement with the morphology of the molecular gas component reported in the velocity range from 106.6 to 116.6 \kms\ toward higher Galactic longitudes, assumed to be related with the PDR (see Sect. \ref{molec}). 
It is important to mention, however, that an inspection of all the \hi\ emission channels show that there is no emission  that could be interpreted as the approaching and receding caps. 
 This absence may be indicating either that the structure has a ring morphology or, as pointed out by \citet{caz05}, that there is significant velocity dispersion, making the receding and/or expanding caps  hard to detect.

\subsection{Infrared}\label{s-ir}

\begin{figure}
\centering
\includegraphics[width=9cm]{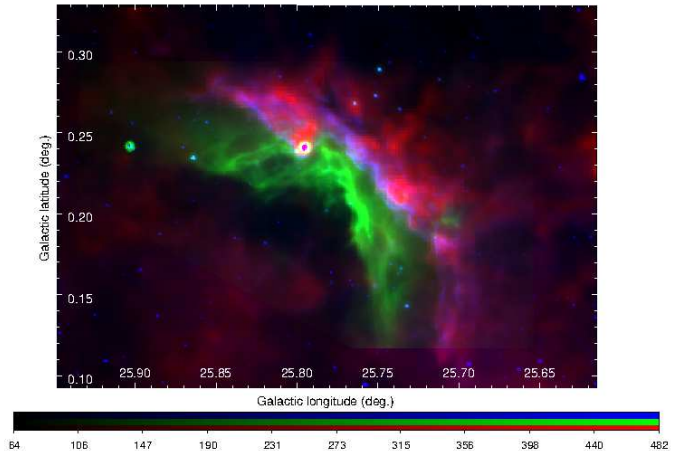}
\caption{RGB image of \g\, and surroundings. Red: emission at 250  $\mu$m (Herschel); green: emission at 24 $\mu$m (Spitzer); and blue: emission at 8 $\mu$m (Spitzer).}
\label{ir-2}
\end{figure}

At infrared wavelengths, the region appears to be very conspicuous, as shown in Fig. \ref{ir-2}. In fact, 
\g\, was also identified in the WISE images and catalogued  as G025.867+00.118 \citep{and14}. Figure \ref{ir-2} is a composite RGB image of the region, where the emission at 250 $\mu$m, 24 $\mu$m, and 8 $\mu$m is displayed in red, green, and blue respectively. Since the 8 $\mu$m emission mainly originates in the polycyclic aromatic hydrocarbons (PAHs), which are destroyed inside an \hii \ region, it reveals the presence of a photodissociation region (PDR; \citealt{ht97}). The emission at 24 $\mu$m shows the dust heated by the energetic stellar photons while 250 $\mu$m\, data trace the cold dust big grains (BG) emission.
 
A remarkably strong source is readily observed at ($l$, $b$) = (25\fdg8, 0\fdg24) in Fig. \ref{ir-2}. This source, listed as G025.7961+00.2403 in the MSX catalog, is classified as a compact \hii\, region based on its infrared colors. We will discuss its nature and possible origin in Section \ref{sf}.

\subsubsection{Dust temperature maps}

Dust temperatures in the region were estimated using the emission at 70 and 160 $\mu$m.
In a far infrared (FIR) image, the dust temperature ($T_d$) in each pixel can be obtained assuming that the dust in a single beam is isothermal and that the observed ratio of 70 to 160 $\mu$m is due to blackbody radiation from dust grains at $T_d$ modified by a power law emissivity. The method used to calculate the dust color temperature is the one used by \cite{sch05}.
The 70 $\mu$m\, image was convolved to the resolution of the 160 $\mu$m\, image.

The flux density emission at  a wavelength $ \lambda_i$ is given by

\begin{equation}
F_i = \frac{2hc}{\lambda_i^3 (e^{hc/(\lambda_i k T_d)}-1} N_d \alpha \lambda_i^{-\beta} \Omega_i
\end{equation}
where $N_d$ represents the column density of dust grains, $\alpha$ is a constant that relates the flux to the dust optical depth, $\beta$ is the emissivity spectral index, and $\Omega_i$ is the solid angle subtended at $\lambda_i$ by the detector.

\begin{figure}
\centering
\includegraphics[width=9cm]{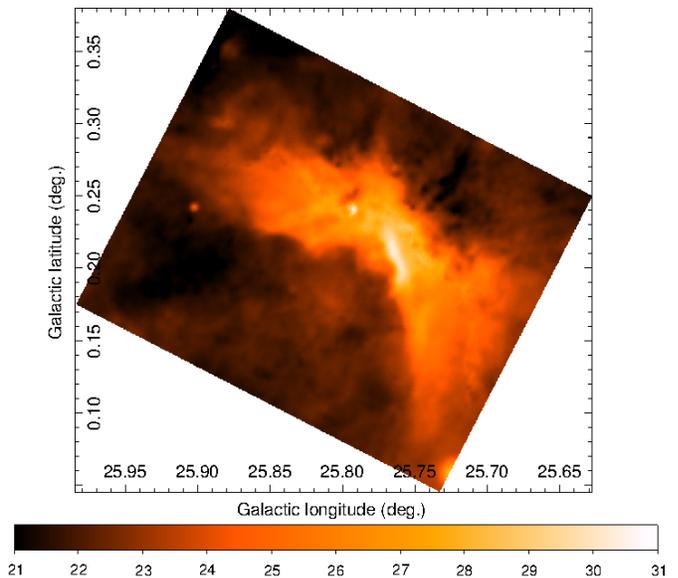}
\caption{ Color temperature map based on emission at 70 and 160 $\mu$m. Color bar units are expressed in K.}
\label{ir-1}
\end{figure}

Assuming that the dust emission is optically thin at 70 and 160 $\mu$m\, and that $\Omega_{70} \sim \Omega_{160}$, and adopting an index of $ \beta = 2$ we can write the ratio, $R$, of the flux densities as
\begin{equation}
R = (0.4)^{-(\beta +3)} \frac{e^{90/T_d}-1}{e^{205/T_d} -1} 
\end{equation}
The dust temperature, $T_d$, is derived from this equation. Fig. \ref{ir-1} shows the spatial distribution of the dust color temperature, which varies from $20$ K to $31$ K. The uncertainties were estimated  to be about $\sim 10-15 \%$. The map shows a clear temperature gradient from the central regions (the inner arc-shaped structure, where the ionizing stars are located) to the periphery of the complex, while low temperatures ($\sim$ 23 K) are measured at the location of the  molecular clumps (see Section \ref{molec}).

\begin{figure}
\centering
\includegraphics[width=7.8cm]{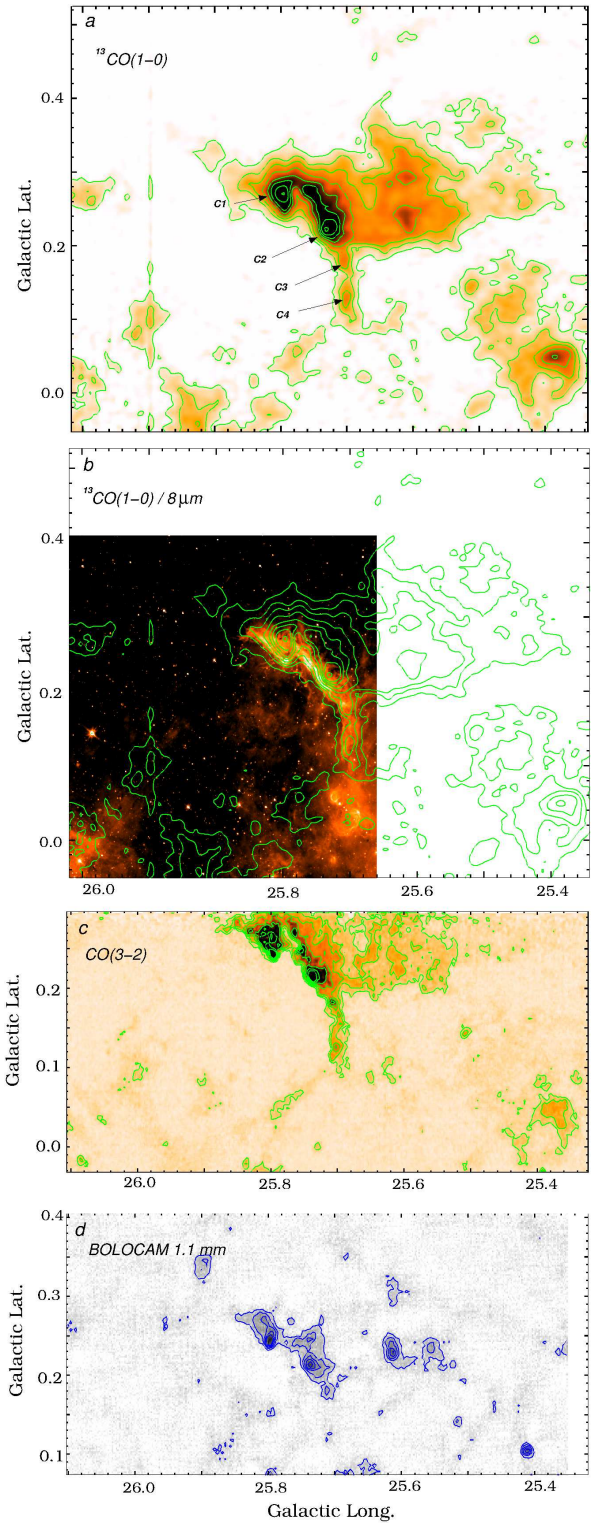}
\caption{{\it Panel a:} $^{13}$CO(1-0) emission map in the velocity range from 106.6 \kms\ to 116.6 \kms. Contour levels start from 0.2 K \kms with a contour spacing of 0.2 K \kms. The positions of molecular clumps C1 and C2 are indicated (see text). {\it Panel b:} Composite image showing the emission of the $^{13}$CO(1-0) line (green contours) over the {\it Spitzer} 8 $\mu$m emission (red colour). {\it Panel c:} CO(3-2) emission map in the velocity range from 106.6 \kms\ to 116.6 \kms. Contour levels start from 1 K \kms\, with a contour spacing of 1 K \kms. {\it Panel d:}  Bolocam 1.1 mm emission. Contour levels start from 100 mJy beam$^{-1}$ ($\sim 3\sigma$) with a contour spacing of 100 mJy beam$^{-1}$.    }  
\label{co}
\end{figure}

\subsection{Millimeter and sub-millimeter emission}\label{molec}

\subsubsection{Carbon monoxide emission}

 In Fig. \ref{co} (panel {\it a}) we show the  \hbox{$^{13}$CO(1-0)} emission distribution map integrated within the velocity interval from 106.6 \kms\ to 116.6 \kms. The molecular  gas at this velocity interval depicts a partially arc-shaped structure, with the strongest emission coincident with the IR emission (see Sect. 3.3). This velocity interval is coincident with the velocity of the RRLs detected in the region (see Sect. 2). In panel {\it b}, we show a composite image of the 8 $\mu$m emission with the \hbox{$^{13}$CO(1-0)} emission. The molecular gas shows an excellent morphological correspondence with the infrared emission, which indicates that this molecular component is being ionized by a stellar source (or sources) at lower Galactic Longitude giving rise to a  PDR. The presence of strong HI emission almost between the radio continuum  and the molecular emission  (see Figs. \ref{hi} and \ref{continuo}) suggests the existence of a central  neutral gas layer between the ionized gas and the molecular gas giving  additional support to the PDR scenario.  The molecular emission shows a sharp arc-shaped cutoff in the direction of the radio continuum emission while the intensity of the emission decreases smoothly toward the center of the molecular cloud. This layout, which is indicative of a density gradient in the molecular distribution,  suggests that the molecular gas has undergone ionization and has probably been  compressed on the front side as a result of the expansion of the ionization front and/or by the action of winds of the powering star/s. The incomplete shell structure of the molecular component and the location of the brightest molecular emission (behind the radio continuum emission at 1420 MHz) give further support to the previously proposed scenario of an \hii\ region density bounded toward higher Galactic longitudes and ionization bounded toward lower Galactic longitudes.  In the panel {\it c} of Fig. \ref{co} we show the CO(3-2) emission distribution in the region, which shows a similar feature.

It is clear from Fig. \ref{co}  that the  molecular emission is not uniformly distributed, and several clumps can be distinguished in the brightest region, next to the ionization front.  From the present data, it is difficult to ascertain whether these clumps  are preexisting concentrations in the molecular structure, or instead they were formed by the accumulation of molecular gas due to the expansion of the ionization front and/or the action of stellar winds. Observational evidence (see  previous paragraph) suggests the latter scenario, although for the case of C1, its cometary shape in the continuum emission at 1.1 mm (see Sect. 3.4.2) might indicate that this clump was previously formed.

The physical properties  of these clumps must be estimated if a study of the  molecular gas is intended. In the following analysis we will concentrate only on  the molecular clumps located along the PDR, which are those very likely formed by the expansion of the ionized gas over the parental molecular cloud. These clumps will be hereafter dubbed as clumps C1, C2, C3, and C4. It is worth to point out that clumps C1 and C2 were identified by \citet{anda09} as the molecular components of U25.80+0.24 and C25.77+0.21, respectively. The location and size of clump C1 is coincident with the MSX source G025.7961+00.2403 and the bright IR source reported at 8, 24, and 250 $\mu$m in Sect.~3.3 (see Fig.\ref{ir-2}).  

\begin{table*}
\caption{Physical properties derived for the CO clumps. } 
\label{coprop}
\centering  
\begin{tabular}{c c c c c c c c c c c c c c}
\hline \hline
  &  A$_{\rm clump}$ & $R_{\rm eff}$   & V$_{\rm LSR}$  & $T^{13}_{\rm peak}$ & $T_{\rm exc}$ & $\tau^{13}$  & $N({\rm H}_2$)  &  $\tau^{13}_{\rm peak}$   &  $N({\rm H}_2$)$_{\rm peak}$ &  $M_{\rm H_2}$ & $n_{\rm H_2}$   &  $N_{\rm int}$(H$_2$) & $M_{\rm int}$    \\
   & (10$^{-6}$ ster)  &  (10$^{19}$ cm)    &  (\kms)   & (K) & (K)  &  & (10$^{22}$ cm$^{-2}$)  &  &  (10$^{22}$ cm$^{-2}$) & (10$^4$ $M_{\odot}$)   &  (10$^{3}$ cm$^{-3}$)   &  (10$^{22}$cm$^{-2}$) & (10$^4$ $M_{\odot}$)\\ 
\hline 
C1 &  0.98       &   1.12   &  110.3    &  7.9   & 46.5  & 0.06 & 3.2  & 0.19   & 9.9   & 2.8   &   2.1 & 2.6 & 0.8 \\
C2 &  2.05       &   1.62   &  110.6    &  8.1  &  32.1 &  0.08 & 2.0  & 0.31   & 7.9   & 3.6   &   0.9 & 2.3 & 1.1 \\
C3 &  0.23       &   0.54       &   111.6     &  4.8  & 24.8  &  0.05 & 0.8  & 0.24   & 3.7 & 0.16  &  1.1  & 1.1 & 0.3  \\
C4 &  0.83       &   1.03       &    111.7    & 4.4   & 16.5 &  0.12 & 0.9  & 0.42   & 3.1 & 0.31  &  0.3  & 0.9 & 0.3 \\
\hline
\end{tabular}
\end{table*}

In Table \ref{coprop} we present some physical properties derived for the CO  clumps. We define the area of each clump ($A_{\rm clump}$, Col. 3) by the contour level corresponding to half the $^{13}$CO peak emission (T$^{13}_{\rm peak}$, Col. 5). The optical depth of the $^{13}$CO emission ($\tau^{13}$, Col. 5) and the $^{13}$CO column density (N($^{13}$CO)) were calculated from
\begin{equation}
\tau^{13} \ =\ -{\rm ln}\ \left( 1\ -\   \frac{T^{13}_{\rm peak}} {5.29\ {J[T_{\rm exc}] - 0.164}}      \right)
\label{eq:tau13}
\end{equation}
and
\begin{equation}
 N(^{13} \rm CO)\ =\ 2.42 \times 10^{14}\ \tau^{13}\ \frac{ \Delta v\  {\it T}_{\rm exc}} {1-e^{(-5.29 / {\it T}_{\rm exc})  }   }\   \ \textrm{(cm$^{-2}$)}
\label{eq:N13CO}
\end{equation}
where $J[T({\rm K})]=1/(e^{5.29/T(\rm{K})}-1)$, $\Delta$v is the FWHM line width, and $T_{\rm exc}$ is the excitation temperature of the J=1$\rightarrow$0 transition of the $^{13}$CO (Col. 6). The excitation temperature is usually  estimated by the peak temperature  of  the optically thick CO(1-0) emission, through the equation
\begin{equation}
 \qquad T_{\rm exc}\ =\ \frac{ 5.53 } {{\rm ln}\ [ 1\ +\  5.53/ (T_{\rm peak}({\rm CO})\ +\ 0.819) ]    }  \quad \quad  \textrm{(K)}
\label{eq:texc}
\end{equation}
{\bf Since no CO(1-0) line data with spatial resolution better than $\sim$ 9$'$ is  available for this quadrant of the Galaxy \citep{dame01},}  we made use of the empirical relation 
\begin{equation}
  T_{\rm peak}({\rm CO(1-0)})\ =\ \frac{T_{\rm peak}({\rm CO(3-2)})\ +\   0.222}{0.816}
\label{eq:oka}
\end{equation}
derived by \citet{oka12} for the Galactic center. Then, the H$_2$ column density (N(H$_2$), Col. 8) was derived using \hbox{N$(\rm H_2)$ / N$(^{13} {\rm CO})$} = \hbox{5 $\times$ 10$^{5}$} \citep{d78}. The peak optical depth ($\tau^{13}_{\rm peak}$) and peak column density (N(H$_2$)$_{\rm peak}$) are indicated in Cols. 9 and 10, respectively. 

The  molecular mass was calculated using
\begin{equation}
\label{eq:masa}
   M(\rm H_2)\ =\  (m_{sun})^{-1}\  \mu\ m_H\ \sum\ A_{clump}\  N{\rm (H_2)}\ {\it d}^2 \quad  \quad \quad   \textrm{(M$_{\odot}$)}
\end{equation}
where  m$_{\rm sun}$ is the solar mass ($\sim$ 2 $\times$ 10$^{33}$ g),    $\mu$ is the mean molecular weight, which is  assumed to be equal to 2.8 to allow for a relative helium abundance of 25\%,  m$_{\rm H}$ is the hydrogen atom mass   ($\sim$ 1.67 $\times$ 10$^{-24}$ g),  and $d$ is the distance (estimated to be 6.5 kpc; Section \ref{dist}).  The volume density  ($n_{\rm H_2}$, Col. 12) was estimated considering a spherical geometry, as
\begin{equation}
n_{\rm H_2}\ =\ \frac{M_{\rm H_2}}{4/3\ \pi\ R_{\rm eff}^3\ \mu\ m_{\rm H} }
\end{equation}
where $R_{\rm eff}$ (Col. 2) is the effective radius of the clump, estimated as $R_{\rm eff}$ = $\sqrt{A_{\rm clump}/ \pi}$

A different approach to estimate the column density can be made using  the CO(3-2) integrated intensity emission. 
\begin{equation}
  N_{\rm int}({\rm H_2})\ =\ X\ \times  \int{\frac{  T({\rm CO(3-2)})\ d{\rm v}}{0.7}}, 
\label{eq:nint}
\end{equation}
where $X$ is an empirical factor that has been shown to be roughly constant for Galactic molecular clouds. For the  the \hbox{$^{12}$CO(1-0)} line, the $X$ value is about 1.9 $\times$ 10$^{20}$ K \kms\ \citep{sm96}. Since we use the integrated intensity emission of the CO(3-2) line, we need to adjust the value of $N_{\rm int}({\rm H_2})$  using a correcting factor of $\sim$  0.7 \citep{oka12}. Hence, the  integrated mass ($M_{\rm int}$) is calculated using Eqs. \ref{eq:masa} and \ref{eq:nint}.

\subsubsection{Continuum 1.1 mm emission}\label{bolocam}

The continuum emission at 1.1 mm is usually dominated by  optically thin thermal emission from cold dust embedded  in dense material (e.g. dense star-forming cores/clumps and filaments), turning this emission into one of the most reliable tracers of dense molecular gas. 

In the panel {\it d} of Fig.~\ref{co} we show the 1.1 continuum emission image obtained from the Bolocam survey. As expected, the emission appears concentrated towards the position of the PDR. Three bright structures can be discerned coincident with the CO clumps C1, C2, and C3. For the sake of clarity, we will also refer to these structures as C1, C2, and C3. C1 and C2 were identified as G025.797+00.245 and G025.737+00.213, respectively in the catalog of Bolocam Galactic Plane Survey (BGPS) sources \citep{shi13}. The presence and location of these submillimeter sources along the border of the \hii\, region  is another confirmation for the existence of high-density molecular gas, which was probably accumulated due to the expansion of the ionization front over the molecular environment. This makes clumps C1, C2, and C3 excellent candidates to search for star formation activity; a special mention deserves C1, whose cometary shape suggests the existence of a pre-existing molecular clump where an ionized boundary layer (IBL) originated by nearby O-type stars may be acting to induce the gravitational collapse.  Triggered star formation scenarios will be further tested on the regions of C1, C2, and C3 (see Sect.~4.2.2).

We estimated the total (H$_2$ + dust)   mass   of  C1, C2, and C3 from their integrated 1.1 mm emission, assuming that the emission is optically thin, and using the equation of \citet{hil83},
\begin{equation}
\qquad M_{(\rm tot)} = R \ \frac{S_{1.1 mm}\ d^2}{\kappa_{1,1 mm}\ B_{1.1 mm}(T_{\rm dust})}
\label{masa}
\end{equation}
where $R$ is the gas-to-dust ratio, assumed to be 186  \citep{dra07}, $S_{1.1 mm}$ is the flux density, $d$ is the distance (6.5 kpc), $\kappa_{1.1 mm}$ is the dust opacity per unit mass at 1.1 mm assumed to be 1.0 cm$^2$ g$^{-1}$  (estimated for dust grains with thin ice mantles in cold clumps; \citealt{osse94}), and $B_{1.1 mm}(T_{\rm dust})$ is the Planck function for a temperature $T_{\rm dust}$. 
The beam-averaged column density ($N_{\rm H_2}$) of sources C1, C2, and C3 were calculated using
\begin{equation}
\qquad  N_{\rm H_2}\   =\ R\ \ \frac{I_{\rm peak}}{\Omega_{\rm beam}\  \kappa_{1.1 mm}\ \mu\ m_{\rm H}\  B_{1.1 mm}(T_{\rm dust})}.
\label{cmd}
\end{equation}
 where $I_{\rm peak}$ is the 1.1 mm continuum emission peak intensity, $\Omega_{\rm beam}$ is the beam solid angle \hbox{($\pi$ $\theta_{\rm HPBW}^2$ / 4\ ln(2))},  $\mu$ is the mean molecular weight (assumed to be 2.8 considering a relative helium abundance of 25\%), and $m_{\rm H}$ is the mass of the hydrogen atom. 

The estimated masses, column densities, and dust temperature used for the calculations (obtained from Fig.~\ref{ir-1}),  are listed in Table  \ref{mass-bolocam}.  A direct comparison between Tables \ref{coprop} and \ref{mass-bolocam} shows a good agreement in masses and densities for C1 and  C2. A discrepancy, nonetheless, is observed in the estimated masses for  C3, very likely due to uncertainties in the estimation of the clump's boundaries.

\begin{table}
\caption{Parameters derived from the Bolocam 1.1 mm emission.} 
\label{mass-bolocam}
\centering  
\begin{tabular}{c c c c} 
\hline \hline
 & $T_{\rm dust}$ &  $M_{(\rm tot)}$ & $N_{\rm H_2}$ \\
  & (K) &  (10$^4$ M$_{\odot}$) & $ (10^{22}$cm$^{-2}$ ) \\ 
\hline 
C1 & 25.4 &2.8 & 3.5  \\
C2 & 24.0 &3.4  & 2.6\\
C3 & 25.6 &0.02 &  1.1\\
\hline
\end{tabular}
\end{table}

\section{Discussion}

\subsection{Origin of the structure}\label{orig}

In this Subsection we attempt to find the stars responsible of the origin of \g. 
Considering that early-type high-mass stars have a major importance in perturbing the interstellar medium, we performed a search for such stars in the whole area by inspecting every catalog available, but  we obtained negative results.
The paucity of candidate stars can be explained by the high absorption produced by the gas and dust column intervening up to the distance of 6.5 kpc, which can turn even bright early-type stars hard to detect.  
Taking into account that the extinction in the visual band is on the order of 1.8 magnitudes per kiloparsec near the Galactic plane \citep{whi03}, the visual absorption for a distance of 6.5 $\pm$ 1.0 kpc would be between 10 and 13.5 mag.

\begin{figure}
\centering
\includegraphics[width=9cm]{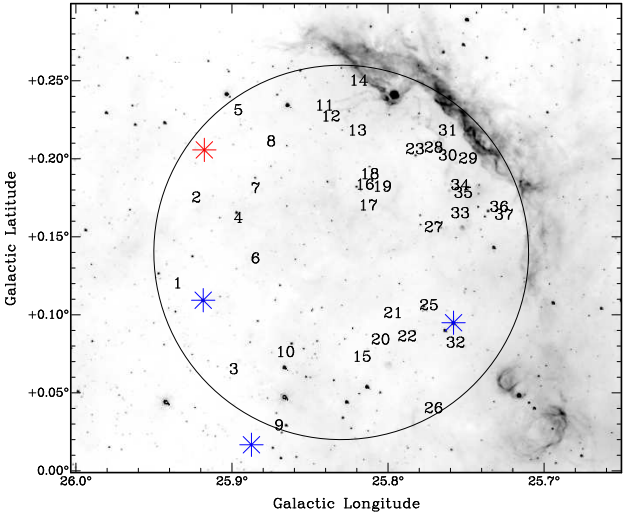}
\caption{ Spitzer emission at 8 $\mu$m. The circle shows the region considered in the search for the exciting stars. The red asterisk indicates the location of the WR star HDM\,40, while 
the   numbers correspond to the ID number of the 2MASS sources given in the first column of Table \ref{estrellas}. 
 The blue asterisks show the location of the three RSGs present in the region. }
\label{reg}
\end{figure}

Another estimation for the visual absorption A$_V$ can be inferred using the Galactic extinction model developed by \citet{che99}. For positive Galactic latitudes,they propose the relation
$$\frac{E(B-V)(D, l, b)}{E(B-V)(\infty, l, b)}= 1\, -\, \exp\,(-\frac{D\, \sin b}{ h})$$
where $E(B-V)(\infty, l, b)$ is the total reddening in the line of sight ($D =  \infty$), and $ h $ is the scale height of the Galactic Plane absorbing dust, $h = 117.7 \pm 4.7$ pc \citep{kos14}. 
To apply this procedure we first inspected the infrared emission distribution  in the region of \g\, to pick up the region where the exciting stars are more probably located.
The \hii\, region has an arc-like shape, hence the exciting stars are expected  to appear near its curvature center.
In Fig. \ref{reg}   the region used to look for the ionizing stars is indicated.
The value of $E(V-B)(\infty, l, b)$ 
was obtained using the dust map of the Galaxy constructed by \citet{sch11}. 
Assuming $A_V = 3.1\, E(B-V)$, we estimated that the visual absorption in the region is in the range between 10 and 16 mag. 
This high extinction can explain why  the ionized region,  being very bright in the radio continuum, is not visible at optical wavelengths, and implies that any candidate early-type ionizing star/s must be searched for in infrared wavelengths.

Several Wolf-Rayet (WR) stars have been recently identified in the infrared by \citet{mau11}. Among them, there is one  located  at ($l, b$) = (25\fdg92, 0\fdg21) (red asterisk in Fig.\ref{reg}), which could be related to \g\,.
This object is the 2MASS source 18375149-0608417 and was identified as HDM\,40 and classified as a WC9d star, since strong evidence has been found of thermal dust emission associated.
\citet{mau11}  estimated  for HDM\,40 a photometric distance of 4.9 kpc, with an uncertainty probably up to 25\% -- 40\%. The authors claim, however, that this distance should be taken with caution, since the probable thermal dust emission from this star can make the adopted colors and K$_s$- band absolute photometry not reliable. 

Red supergiants can be used to trace recent high-mass star formation even in heavily obscured regions. These high-mass stars are young \citep[between 8 and $25\:$Ma;][]{eks2013} and very luminous (from $10^{4.5}$ to $10^{5.8}\:L_{\odot}$). However, unlike ionizing high-mass stars, RSGs present late spectral types (K and, mostly, M) and red colours. Therefore, they are significantly brighter in infrared bands, and they are more accessible targets for spectroscopic surveys in such bands. In consequence, RSGs are a powerful tool to study highly-extinguished high-mass populations. There are two large high-mass clusters close to \g\,: RSGC1 \citep[$l=25\fdg27\:$, $b=-0\fdg16\:$;][]{fig06,dav08} and RSGC2 \citep[$l=26\fdg19\:$, $b=-0\fdg07\:$;][]{dav07}. Due to their high extinctions (between 11 and $26\: $mag in $V$), the only stellar components observable in them are RSGs. These two clusters present radial velocities ($123.0\pm1.0\:$\kms\, and $109.3\pm0.7\:$\kms\,, respectively) and thus kinematic distances ($6.6\pm0.9\:$kpc and $5.8^{+1.9}_{-0.8}\:$kpc, respectively) compatible with \g. Given their high initial masses ($30\pm10\:$kM$_{\odot}$ and $40\pm10\:$kM$_{\odot}$), estimated through population synthesis models from their current RSG population, they should have hosted a significant number of O stars. However, given that the clusters have ages of $12\pm2\:$Ma and $17\pm 3\:$Ma, largely exceeding typical lifetimes of O-type stars, these latter are not expected to exist currently in the clusters. Thus, these clusters are necessarily older than \g\,, where a significant number of O~stars is required to explain its radio continuum emission. 

\begin{figure*}
\centering
\includegraphics[width=14cm]{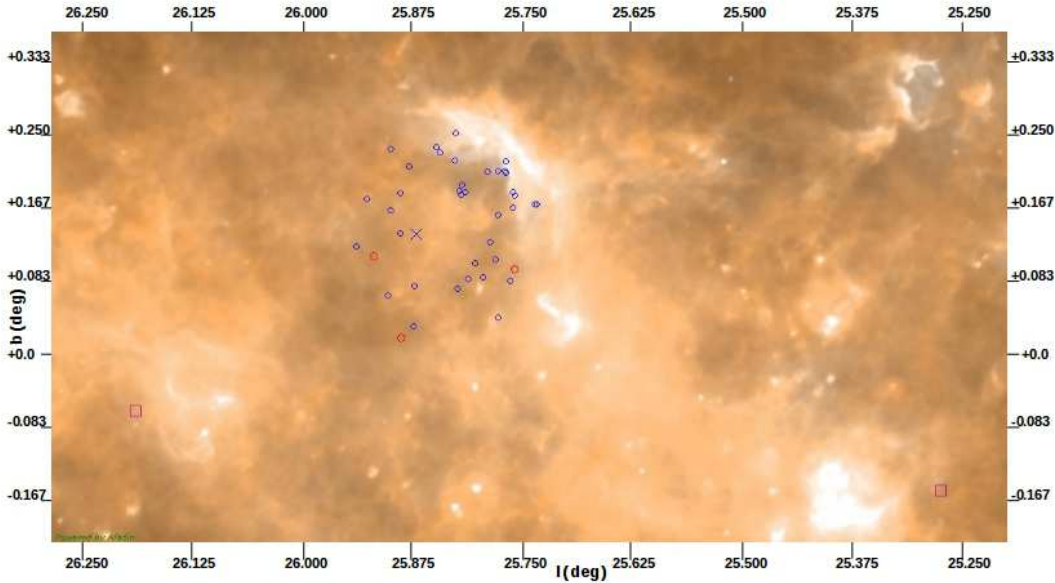}
\caption{ Two colour composite image of \g\, and surroundings. Blue: PACS 70 $\mu$m, Red: PACS 160  $\mu$m. The OV source candidates and the RSG known in the region are represented by blue and red circles, respectively. The blue cross indicates the location of the center of \g\,, and the squares the position of the RSG clusters, RSG1 and RSG2.
  }
\label{cumulos}
\end{figure*}

Despite the difference in age between the clusters and \g\,, it is not unlikely that \g\, may be related somehow to these clusters. It has been determined that these clusters are part of a extended stellar association \citep{neg12}, as there is an over-density of RSGs around them having compatible radial velocities, which is indicative of intense high-mass star forming activity in the region. In fact, the center of \g\, is located between RSGC1 and RSGC2 (see Figure~\ref{cumulos}), separated of them by 39\farcm8 and 22\farcm6, respectively. At the estimated distance of \g\,, $6.5\pm1\:$kpc, these separations correspond to a \textit{minimum} physical distances of $75^{+12}_{-11}$ and $43^{+6}_{-7}\:$pc. Moreover, \g\, has a radius of about 8\farcm5 ($16\:$pc). Therefore, the edges of \g\, are even closer to these clusters. Considering that both clusters are separated by 54\farcm6 ($103\:$pc at the distance of \g\,) and since they seem to belong to the same association, it is unlikely that a high-mass star forming region as \g\, were not related to them and to the over-density of RSGCs around them. Moreover, the age difference between the clusters ($5\:$Ma) is not much shorter than the difference that we should expect between RSGC1 and the O-stars in \g\, ($8 - 9\:$Ma). Therefore, we suggest that \g\, may represent the youngest star-forming burst of this high-mass star association. 

According to evolutionary models \citep{eks2013}, only those RSGs of higher masses and luminosities come from the evolution of O~stars. Thus, the presence of RSGs related to \g\, may provide hints about its past O-star population history. Given the over-density of RSGs in this region, we searched for those lying close to \g\, in the spectroscopic catalogue of Dorda et al. (in preparation). This catalogue provides spectral classifications as well as radial velocities for the stars observed \citep{dor16}. We found three RSGs in the area of \g\, with compatible radial velocities (see Table~\ref{rsgs} and Fig. \ref{reg}). We used their spectral types together with the work of \cite{lev05} to estimate their $(J-K_{\rm s})_{0}$. Then we calculated $E(J-K_{\rm s})$ using the 2MASS photometry and we dereddened $k_{\rm s}$ applying the calibration of \citet{rie85}. We also calculated the bolometric magnitude, using the bolometric corrections proposed by \cite{lev05}. Finally, considering that we want to evaluate if these RSGs can be related to \g, we used the distance to this \hii\, region to calculate their absolute magnitudes. The masses and corresponding age interval of the RSGs was roughly estimated through $M_{\rm bol}$, since evolutionary models predict that the luminosity of a RSG depends strongly on its initial mass. For this estimation we used the evolutionary tracks of Geneva~\citep{eks2012}. Under the distance assumed, the RSGs \#2 and \#3 are too faint to have evolved from O~stars, and thus, they are too old to be related in any way with \g. However, for the RSG \#1, with a $M_{\rm bol}\sim-8$, we estimate a $M_{\rm i}\sim20\:$M$_{\odot}$ which may well imply a late O~star origin (O8V--O9V). A RSG having such mass should have an age between 8 and $9\:$Ma, which implies that all but the latest O stars have already died if coeval, and thus it should be related to a population whose most massive members at present are O9V stars. In consequence, to confirm the relation between the RSG \#1 and \g\, it would be necessary to know the spectral types and radial velocities of the early-type high-mass stars in this region.

\begin{table*}
\caption{Red supergiants from \citet{dor16} in the area of \g\, and with compatible radial velocities.  } 
\label{rsgs}
\centering  
\begin{tabular}{l c c c c c c c }
\hline \hline
\# & 2MASS source & $l$ (deg.) & $b$ (deg.) & $V_{\rm LSR}\pm1$ (\kms\,) $^{\rm a}$& Spectral type $^{\rm a}$& $K\:$(mag) $^{\rm b}$& $M_{\rm bol}$ (mag) $^{\rm b}$\\
\hline 
1 & 18375756-0620155 & 025.75802 & +00.09493 & 97 & M2.5 & -10.77 & -7.94\\
2 & 18382868-0615304 & 025.88752 & +00.01671 & 103 & M3.5 & -8.15 & -5.27\\
3 & 18381223-0611186 & 025.91842 & +00.10935 & 101 & M2 & -10.26 & -7.46\\
\hline
\end{tabular}
\begin{list}{}{}
\item{$^{\rm a}$} Spectral type and $V_{\rm LSR}$ are averaged values from the multiple epochs available.
\item{$^{\rm b}$} The absolute magnitudes were calculated assuming the distance of \g\,: $6.5\:$kpc.
\end{list}
\end{table*}

Bearing in mind that early-type high-mass stars have a huge impact onto their surrounding gas via their high rate of energetic photons and stellar winds, the presence of cavities, shells, and \hii\, regions can be used as good tracers for their existence. Thus, as mentioned above, taking into account the morphology of \g, we constrained the region where the exciting stars are most probably located (see Fig. \ref{reg}) and looked for the exciting star candidates using the 2MASS source catalog \citep{cut03}. 
We selected the candidates using the infrared reddening-free pseudo color $Q_{IR}= (J-H) - 1.83 (H-K_{\rm s})$ , picking those  2MASS sources with $-0.15 < Q_{IR} < 0.1$, indicative of main-sequence stars  \citep{com02}. 

\begin{figure}
\centering
\includegraphics[width=9cm]{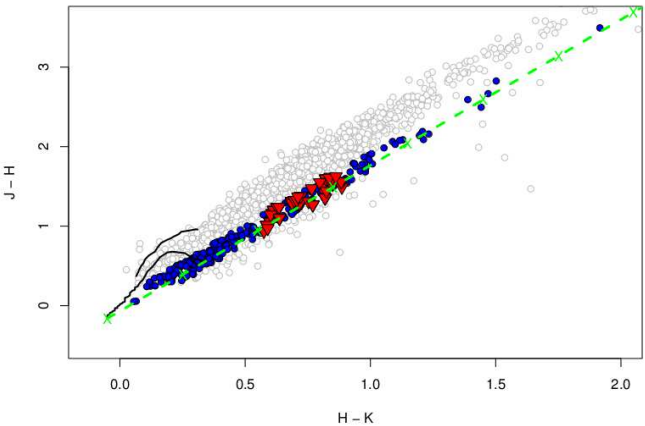}
\caption{ Color-color diagram of the 2MASS sources projected onto the region shown in Fig. \ref{reg}.  Main-sequence candidates are plotted with blue circles.  The positions of the dereddened early-type main sequence and giant stars are
shown by black lines. The reddening vector for  O9 V stars (dashed green line) is indicated. The OV source candidates to be related to \g\, are represented by red triangles.
  }
\label{cc}
\end{figure}

We found 1764 sources located  inside the region having the best photo-metric quality (ph-qual = AAA) in the three bands  $J$ (1.235 $\mu$m), $H$ (1.662 $\mu$m), and $K_{\rm s}$ (2.159 $\mu$m). Among them, 262 are classified as main-sequence star candidates. Figure \ref{cc} shows a color-color diagram depicting the distribution of  all sources, with the  main-sequence ones plotted in blue. 
The figure also shows the positions of the dereddened  main sequence and giant stars. 
We located the O-stars into the main sequence using the values given by \citet{mar06}, and those given by \citet{tok00} and \citet{dri00} for late-type stars (B to M). The reddening vector for an early type (O9 V) star  
is represented by a dashed green line  using extinction values from \citet {rie85}.
From Fig. \ref{cc} we selected the main-sequence candidates lying along the reddening vector between A$_V$ = 10 and 16 magnitudes and 
computed their  $M_J$, $M_H$, and $M_K$ absolute magnitudes, assuming a distance of $D= 6.5 \pm 1.0 $ kpc.
We then analyzed whether the estimated magnitudes are in agreement with the absolute magnitudes of O-type stars as given by \citet{mar06}. From this comparison we finally found that 37 out of the 1764 2MASS sources could be O-type stars related to \g. They are listed in Table \ref{estrellas}, indicated by red triangles in Fig. \ref{cc} and by white numbers in Fig. \ref{reg}.

 An inspection of their estimated absolute magnitudes (see Columns 8, 9, and 10 of Table \ref{estrellas}) shows that several of them could be brighter than an O9V-type star, for which $M_J = -3.48$, $M_H = -3.38$, and $M_K = - 3.28$ magnitudes \citep{mar06}, thus suggesting that more than one episode of massive stellar formation has taken place along the last million years.

It is important to note that these 37 sources are just candidates of being responsible of the ionized region,
and to confirm this association it is necessary to analyze their spectra. 
Besides, given the several assumptions involved in the method used to select the O-type star candidates,  a certain degree of contamination is to be expected. To estimate the possible contamination level, we applied the same method in a nearby region, covering the same area but centered at ($l$, $b$) (26\fdg0, 0\fdg295). We found that 19 out of 1464 2MASS sources (with AAA quality) are OV star candidates (at a distance of 6.5 kpc and with $A_V$ between 10 and 16 mag), which is about half the number we found in the \g\, region, suggesting that the unrelated objects are probably less than 19.

As mentioned in Section \ref{seccont}, the number of ionizing photons needed to keep the region ionized is very high, $ N_{\rm UV} = (5.0 \pm 1.6) \times 10^{49}$ s$^{-1}$. Such amount of UV photons could be supplied by  several early-type stars, like 3 O5V, 12 O7V, or 63 O9V stars \citep{mar05}.
Moreover, since in the estimation of $ N_{\rm UV}$ the effect of the dust is not taken into account, these figures are just lower limits. In this context, we conclude that most of the 37 2MASS sources found as O-type star candidates could be indeed related to \g. 
Finally, the presence of at least one supergiant suggests that there have been several SN in the area. The absence of catalogued SNRs within this region could be explained either if the explosions ocurred $\gtrsim 10^5$ years ago and the remnants have dissipated into the ISM, or the putative SNRs are confused with line-of-sight emission from other sources along this complicated direction of the Galaxy, towards the Galactic Plane. However, although no detected directly, shock waves driven by past SN explosions must have contributed to shape the ISM and could explain the asymmetry of the shell displayed by \g.

\begin{table*}
\caption{Exciting star candidates parameters. } 
\label{estrellas}
\centering  
\begin{tabular}{l c c c c c c c c c }
\hline \hline
\# & 2MASS source & $l$ (deg.) & $b$ (deg.) & $J$ (mag)& $H$ (mag)& $K_{\rm s}$ (mag) &$M_J^{(*)}$ (mag) & $M_H^{(*)}$ (mag)& $M_K^{(*)}$ (mag) \\
\hline 
1	& 18381209-0609575& 	25.938& 	0.12& 	15.087& 	13.497	& 12.662 &$-1.78/ -3.46$&$-2.27/-3.29$&$-2.50/-3.16$\\
2	& 18375910-0609061& 	25.926& 	0.175& 	14.589	& 13.298& 	12.611& $-2.28/	-3.96$&$-2.47/-3.49$&$-2.55/-3.21$\\
3	& 18381995-0613242& 	25.902& 	0.065& 	14.72& 	13.409& 	12.656&$-2.14/-3.82$&$-2.36/-3.38$&$-2.51/-3.17$\\
4	& 18375890-0610528& 	25.899& 	0.162	& 13.817& 	12.803& 	12.217&$-3.05/-4.73$&$-2.96/-3.98$&$-2.95/-3.61$\\
5	& 18374388-0609002	& 25.899& 	0.231& 	14.328& 	13.185	& 12.583&$-2.54	/-4.22$&$-2.58/-3.60$&$-2.58/-3.24$\\
6	& 18380328-0612113& 	25.888& 	0.136& 	13.141& 	11.939& 	11.325&$-3.72/	-5.40$&$-3.83/-4.85$&$-3.84/-4.50$\\
7	& 18375349-0610555& 	25.888& 	0.181& 	13.54& 	12.214& 	11.508&$-3.32/-5.00$&$-3.55/-4.57$&$-3.66/-4.32$\\
8	& 18374590-0610389& 	25.878& 	0.211& 	13.664& 	12.286& 	11.564&$-3.20/-4.88$&$-3.48/-4.50$&$-3.60/-4.26$\\
9	& 18382436-0615561& 	25.873& 	0.029& 	14.888& 	13.318& 	12.429&$-1.98/-3.66$&$-2.45/-3.47$&$-2.74/-3.40$\\
10	& 18381429-0614414& 	25.872& 	0.076& 	14.964& 	13.358& 	12.522&$-1.90/-3.58$&$-2.41/-3.43$&$-2.64/-3.30$\\
11& 	18373768-0611408	& 25.847& 	0.234& 	14.287& 	12.69& 	11.868&$-2.58/-4.26$&$-3.07/-4.09$&$-3.30/-3.96$\\
12& 	18373856-0612053& 	25.843& 	0.227& 	14.89& 	13.341	& 12.543&$-1.97/-3.65$&$-2.42/-3.44$&$-2.62/-3.28$\\
13& 	18373870-0613133& 	25.826& 	0.218	& 13.759& 	12.657& 	12.02&$-3.11/-4.79$&$-3.11/-4.13$&$-3.14/-3.80$\\
14& 	18373172-0612247& 	25.825& 	0.25	& 14.141& 	13.205& 	12.63&$-2.72/-4.40$&$-2.56/-3.58$&$-2.53/-3.19$\\
15	& 18380952-0617237& 	25.823	& 0.073	& 13.026	& 11.741& 	11.029&$-3.84/-5.52$&$-4.02/-5.04$&$-4.14/-4.80$\\
16	& 18374551-0614256& 	25.821	& 0.184& 	13.46& 	12.137& 	11.44&$-3.40/-5.08$&$-3.63/-4.65$&$-3.72/-4.38$\\
17& 	18374618-0614401	& 25.819& 	0.18	& 13.48& 	12.24& 	11.605&$-3.38/-5.06$&$-3.52/-4.54$&$-3.56/-4.22$\\
18& 	18374373-0614273	& 25.818& 	0.19	& 14.679	& 13.119& 	12.281&$-2.19/-3.87$&$-2.65/-3.67$&$-2.88/-3.54$\\
19& 	18374508-0614503	& 25.814& 	0.182& 13.949	& 12.593& 	11.774&-$2.92/-4.60$&$-3.17/-4.19$&$-3.39/-4.05$\\
20	& 18380578-0617446	& 25.811&	0.084& 	14.849&	13.225	& 12.363&$-2.02/-3.70$&$-2.54/-3.56$&$-2.80/-3.46$\\
21	& 18380121-0617407	& 25.803	&0.101	& 14.098& 	12.539& 	11.656&$-2.77/-4.45$&$-3.23/-4.25$&$-3.51/-4.17$\\
22& 	18380355-0618358	& 25.794&	0.086& 	14.121& 	13.148& 	12.558&$-2.74/-4.42$&$-2.62/-3.64$&$-2.61/-3.27$\\
23& 	18373718-0615341& 	25.789	& 0.206	& 14.584& 	13.051& 	12.202&$-2.28/-3.96$&$-2.71/-3.73$&$-2.96/-3.62$\\
24& 	18375398-0617554& 	25.786& 	0.126& 	14.725& 	13.284& 12.46&$-2.14/-3.82$&$-2.48/-3.50$&$-2.70/-3.36$\\
25& 	18375763-0618478& 	25.78& 	0.106	& 13.171	& 11.859	& 11.173&$-3.69/-5.37$&$-3.91/-4.93$&$-3.99/-4.65$\\
26& 	18381156-0620453& 	25.777& 	0.04	& 14.281	& 13.011& 	12.241&$-2.58/-4.26$&$-2.75/-3.77$&$-2.92/-3.58$\\
27& 	18374651-0617334& 	25.777& 	0.156	& 14.933	& 13.454& 	12.646&$-1.93/-3.61$&$-2.31/-3.33$&$-2.52/-3.18$\\
28& 	18373564-0616112& 	25.777& 	0.207	& 14.042& 	12.755& 	12.037&$-2.82/-4.50$&$-3.01/-4.03$&$-3.13/-3.79$\\
29& 	18373488-0616374& 	25.769& 	0.206	& 15.011& 	13.52& 	12.634&$-1.85/-3.53$&$-2.24/-3.26$&$-2.53/-3.19$\\
30& 	18373506-0616423& 	25.768	& 0.205& 	15.061& 	13.442	& 12.602&$-1.80/-3.48$&$-2.32/-3.34$&$-2.56/-3.22$\\
31& 	18373224-0616205& 	25.768& 	0.218& 	14.87& 	13.396& 	12.63&$-1.99/-3.67$&$-2.37/-3.39$&$-2.53/-3.19$\\
32& 	18380088-0620213& 	25.763& 	0.082	& 14.748& 	13.379& 	12.668&$-2.12/-3.80$&$-2.39/-3.41$&$-2.50/-3.16$\\
33	& 18374279-0618120& 	25.76& 	0.165& 	14.225& 	12.926& 	12.202&$-2.64/-4.32$&$-2.84/-3.86$&$-2.96/-3.62$\\
34	& 18373904-0617427& 	25.76& 	0.183& 	13.459& 	12.405& 	11.805&$-3.41/-5.09$&$-3.36/-4.38$&$-3.36/-4.02$\\
35	& 18373971-0617559& 	25.758& 	0.179& 	14.253&  	13.015& 	12.388&$-2.61/-4.29$&$-2.75/-3.77$&$-2.78/-3.44$\\
36	& 18373913-0619269& 	25.735& 	0.169& 	12.574& 	11.413& 	10.804&$-4.29/-5.97$&$-4.35/-5.37$&$-4.36/-5.02$\\
37 & 	18373896-0619328 & 	25.733& 	0.169& 	14.959& 	13.467& 	12.662&$-1.91/-3.59$&$-2.30/-3.32$ & $-2.50/-3.16$\\

\hline
\end{tabular}
\begin{list}{}{}
\item{*}
Absolute magnitudes estimated for D = 6.5 kpc and $A_v = 10/16$ mag.
\end{list}
\end{table*}

\subsection{Star formation activity in the region}\label{sf}

In what follows, we will study if recent star formation activity  has taken place in the vicinity of \g\, to disentangle whether seeming expanding motions of the \hii\ region could have triggered star formation.
For this purpose, we will try to detect all Young Stellar Object candidates (cYSOs) around the region and analyze their position with respect to the ionized gas and to the molecular condensations.

\subsubsection{Identification of Young Stellar Object candidates}

Primary tracers of stellar formation activity were searched for using  
the MSX Infrared Point Source Catalogue \citep{ega03}, the  WISE All-Sky Source Catalogue \citep{wri10}, and the GLIMPSE point source catalog  \citep{ben03}. MSX sources were selected if their variability and reliability flags were zero and the flux quality Q was above 1 in all four bands. WISE sources with photometric flux uncertainties above 0.2 mag and signal-to-noise ratio lower than 7 in the W1, W2, and W3 bands, were rejected. Finally, Spitzer sources were kept if their photometric uncertainties were lower than 0.2 mag in all four IRAC bands.

Within a 9-arcmin radius circle centered at ($l$, $ b$)\,=\,(25\fdg83, 0\fdg18), a total of  3 MSX, 511 WISE, and 1716 Spitzer sources have been found fulfilling the selection criteria above. To identify the cYSOs among these sources, we adopted the classification scheme described in  \citet{lum02}, \citet{koe12}, and \citet{gut09} for the MSX, WISE, and IRAC data, respectively. Several sources were found to qualify for cYSOs, and are listed in Table \ref{ysos}. The infrared colors of MSX sources permit  to discern between high-mass young stellar object (MYSO) candidates and compact \hii\, region (CHII) candidates \citep{lum02}. Two of the MSX sources selected in this region belong to this latter class.

Before attempting to identify the cYSOs from the listed WISE and Spitzer sources, we selected the non-YSO  sources with excess infrared emission, such as PAH-emitting galaxies, broad-line active galactic nuclei (AGNs), unresolved knots of shock emission, and PAH-emission features. 
A total of 67 and 132 WISE and Spitzer sources, respectively, were dropped from the lists.
Among the remaining 444 WISE and 1584 Spitzer sources, 13 (4 WISE and 9 Spitzer) were identified as Class I sources (i.e. sources where the IR emission arises mainly from a dense
infalling envelope, including flat spectrum objects) and 63 ( 56 WISE and 16 Spitzer) as Class II sources (i.e. pre-main-sequence stars with optically thick disks).

In the case of the WISE sources identified as cYSOs, we discarded those not compatible with T Tauri star candidates. For this purpose, we checked if the 56 sources previously classified Class II stars with photometric errors lower than 0.2 in WISE band 4 had blue colors in excess, {\bf i.e. if their W1, W3, and W4 magnitudes satisfy that $W1 - W3 \leq -1.7 (W3 - W4) +4.3$ \citep{koe12}.}  Among the 56 Class II sources found, we have rejected 9 based on this last criterion. 
On the other hand, protostellar objects with intermediate/high masses can be identified among Class I sources by additionally requiring their band 3 (12 $\mu$m) magnitude to be less than 5 \citep{hig13}. All four Class I WISE sources detected seem to be high-mass protostars, since all of them satisfy this criterion. 
 
\begin{figure}
\includegraphics[width=9cm]{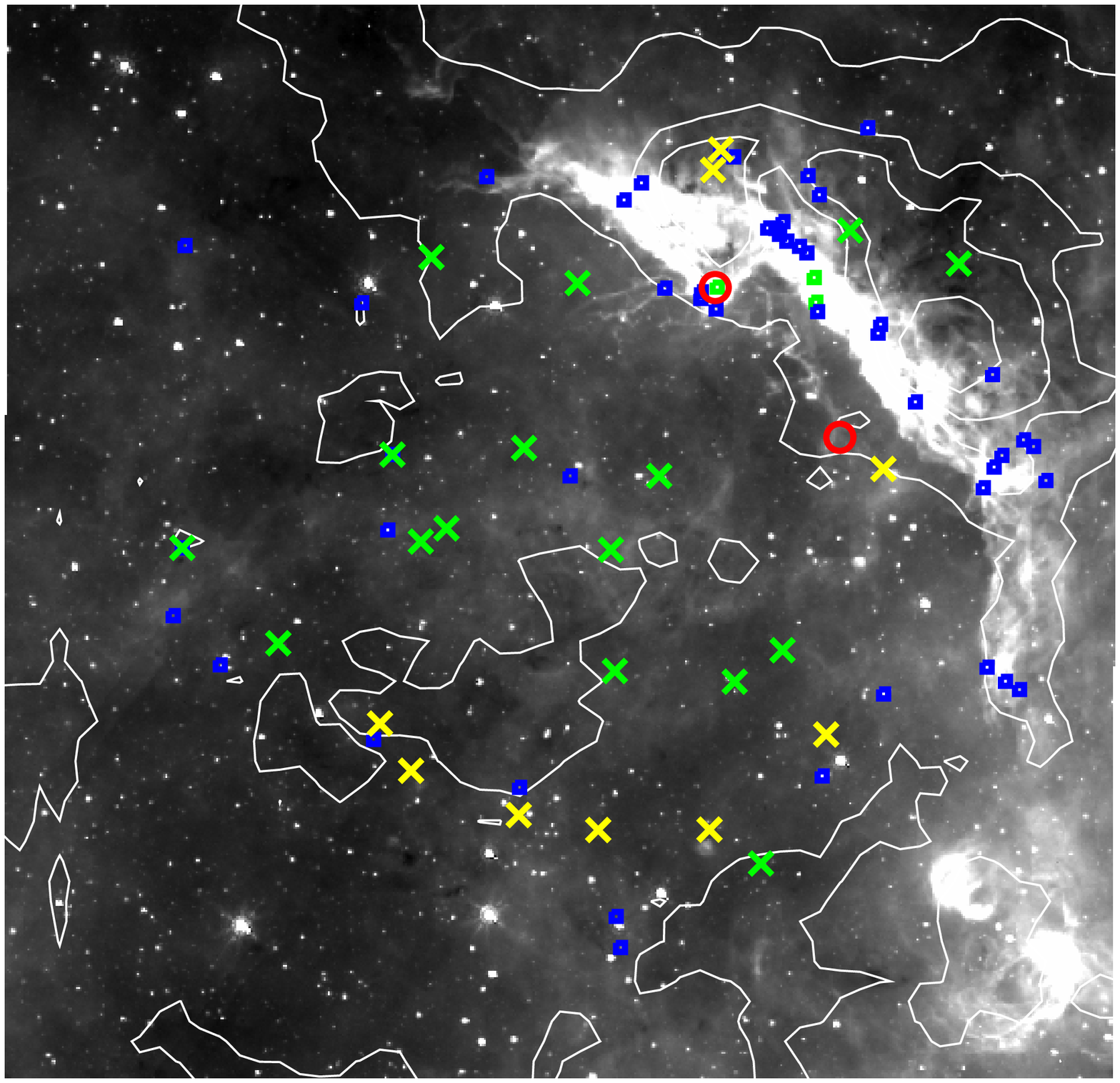}
\caption{ Spitzer 8 $\mu$m  emission distribution showing the PDR.  Contour levels at  0.3, 0.8, 1.0 and 1.3 K delineate the average $^{13}$CO emission between 100 and 115 \kms. The  red  circles indicate the location of the MSX sources. Green and blue squares correspond to WISE Class I and II cYSO , respectively. Yellow and green crosses indicate Spitzer Class I and II sources, respectively.}
\label{fig-ysos}
\end{figure}

The cYSOs remaining in the final list are indicated in Fig. \ref{fig-ysos} and their fluxes/magnitudes are listed in Table \ref{ysos}.

\begin{table*}
\caption{  IRAS, MSX, WISE and Spitzer sources  found to be YSO candidates. \label{ysos}}
\begin{tabular}{c c c c c c c c}
\hline\hline
\multicolumn{8}{c}{{\bf MSX sources}} \\
 $\#$ &  Designation &  {(\it l, b})&  F$_{8}$[Jy] (Q$_{8}$) &  F$_{12}$[Jy] (Q$_{12}$) &  F$_{14}$[Jy] (Q$_{14}$) &  F$_{21}$[Jy] (Q$_{21}$) & Notes \\
\hline
1&G025.7961+00.2403& 25\fdg796, 0\fdg2403&6.8488 (4)&   8.4698 (4)&   5.897 (4) & 25.377 (4)   & CHII, C1\\
2&G025.7597+00.1933&25\fdg7597, 0\fdg193& 0.1439 (3)& 1.304 (3)&      1.0198 (2)   &    5.11 (4) & CHII\\
\hline
\multicolumn{8}{c}{{\bf WISE sources}} \\
$\#$ &  Designation &  {(\it l, b})& W1 [mag]  &    W2 [mag]   &  W3 [mag]  &  W4 [mag] &Notes \\
\hline
3&J183729.32-061225.2&25\fdg820, 0\fdg258 &10.696  $\pm$   0.079 &9.186      $\pm$ 0.035&4.628 $\pm$ 0.021   &2.730  $\pm$  0.050& Class I, C1\\
4& J183726.58-061545.7&25\fdg765, 0\fdg243&  10.435 $\pm$ 0.074 & 9.416 $\pm$  0.061 & 4.498   $\pm$ 0.107 &  0.537  $\pm$ 0.020 &Class I, C2\\
5&J183728.15-061559.8& 25\fdg765,   0\fdg235&   10.330   $\pm$    0.071   &  9.329   $\pm$ 0.060  & 4.362 $\pm$  0.094 &  1.123   $\pm$ 0.038 & Class I, C2\\
6&J183730.52-061415.2& 25\fdg795, 0\fdg240&   7.550   $\pm$   0.021  &       6.522  $\pm$    0.019   &  1.219 $\pm$  0.073 & -2.044  $\pm$  0.016 &Class I, C1\\
7&J183720.07-061447.9	&25\fdg767,	0\fdg274	&9.392  $\pm$ 	0.026&	8.739	  $\pm$ 0.023	&7.279  $\pm$ 	0.039&	4.088	  $\pm$ 0.042&Class II, C2\\
8&J183732.00-061425.2	& 25\fdg795, 0\fdg233	&10.189 $\pm$ 	0.04	&9.83  $\pm$ 	0.046&	5.602  $\pm$ 	0.045&	0.524  $\pm$ 	0.022&Class II, C1\\
9&J183730.15-062034.0&	25\fdg701,	0\fdg193&	11.04  $\pm$ 	0.058	&10.692	  $\pm$ 0.063&	6.218  $\pm$ 	0.065&	1.223  $\pm$ 	0.030&Class II, C3\\
10&J183726.91-061930.3&	25\fdg710,	0\fdg213&	11.615  $\pm$ 	0.14&	11.058  $\pm$ 	0.146&	6.721  $\pm$ 	0.053&	3.319  $\pm$ 	0.064&Class II, C2\\
11&J183731.83-061404.8&	25\fdg800,	0\fdg236&	9.077  $\pm$ 	0.029&	8.52  $\pm$ 	0.029	&4.707	  $\pm$ 0.028&	0.658  $\pm$ 	0.055&Class II, C1\\
12&J183721.37-061325.2&	25\fdg790,	0\fdg280	&9.519	  $\pm$ 0.028&	8.92  $\pm$ 	0.028&	7.497  $\pm$ 	0.075&	3.862  $\pm$ 	0.029&Class II, C1\\
13&J183731.37-061402.1&25\fdg800,	0\fdg238&	8.16  $\pm$ 	0.024&7.600  $\pm$ 	0.024&	3.877  $\pm$ 	0.086&	-0.978	  $\pm$ 0.030&Class II, C1\\
14&J183734.68-062018.6&	25\fdg713,	0\fdg178&	9.999  $\pm$ 	0.035	&9.579	  $\pm$ 0.046	&4.707  $\pm$ 	0.026&	2.802  $\pm$ 	0.036&Class II, C3\\
15&J183746.69-062218.8&	25\fdg706, 0\fdg119&	9.345  $\pm$ 	0.029	&9.029  $\pm$ 	0.027&	5.453  $\pm$ 	0.025&	3.673  $\pm$ 	0.061&Class II, C4\\
16&J183724.91-061434.8&	25\fdg779,	0\fdg258	&10.15  $\pm$ 	0.055&	9.744  $\pm$ 	0.069&	5.28  $\pm$ 	0.028&	1.853  $\pm$ 	0.023&Class II, C1\\
17&J183725.21-061526.0	&25\fdg767	0\fdg250&	9.638  $\pm$ 	0.034&	9.225	  $\pm$ 0.04&	5.229	  $\pm$ 0.05&	1.53  $\pm$ 	0.027&Class II, C2\\
18&J183732.37-061324.2&	25\fdg811,	0\fdg240&	10.864  $\pm$ 0.057&	10.4  $\pm$ 	0.073	&6.883  $\pm$ 	0.121&	1.286  $\pm$ 	0.069&Class II, C1\\
19&J183746.38-062153.6	&25\fdg712,	0\fdg123&	8.308	  $\pm$ 0.023&	7.854  $\pm$ 	0.023&	6.091  $\pm$ 	0.034&	4.188  $\pm$ 	0.072&Class II, C4\\
20&J183728.71-061606.4&	25\fdg764,	0\fdg232&	10.173	  $\pm$ 0.078&	9.589	  $\pm$ 0.053&	4.77  $\pm$ 	0.059&	0.898  $\pm$ 	0.028&Class II, C2\\
21&J183720.95-061508.7&	25\fdg764,	0\fdg268&	9.939  $\pm$ 	0.031&	9.548	  $\pm$ 0.032&	5.264  $\pm$ 	0.022&	1.967  $\pm$ 	0.023&Class II, C2\\
22&J183732.07-062116.5&	25\fdg694,	0\fdg181&	10.792  $\pm$ 	0.05	&9.996  $\pm$ 	0.045&	7.387  $\pm$ 	0.105&	3.731  $\pm$ 	0.054&Class II, C3\\
23&J183727.95-061159.4&	25\fdg824,	0\fdg267	&9.5  $\pm$ 	0.05&	8.816  $\pm$ 	0.032&	4.605  $\pm$ 	0.027&	4.878	  $\pm$ 0.199&Class II, C1\\
24&J183724.91-061449.6&	25\fdg776,	0\fdg256	&10.77  $\pm$ 	0.066&	10.047  $\pm$ 	0.067&	5.189	  $\pm$ 0.027&	2.492	  $\pm$ 0.058&Class II, C1\\
25&J183731.33-061828.3&	25\fdg734,	0\fdg205&	10.262  $\pm$ 	0.037&	9.367  $\pm$ 	0.033&	4.484  $\pm$ 	0.035&	1.442  $\pm$ 	0.043&Class II, C2\\
26&J183746.75-062236.7&	25\fdg702,	0\fdg116&	10.389  $\pm$ 	0.038&	9.933  $\pm$ 	0.035&	6.596  $\pm$ 	0.046&	4.499  $\pm$ 	0.108&Class II, C4\\
27&J183714.88-061522.2&	25\fdg749,	0\fdg289&	5.587  $\pm$ 	0.123&	4.936	  $\pm$ 0.064&	3.359	  $\pm$ 0.017&	2.074	  $\pm$ 0.018&Class II, C2\\
28&J183731.91-062020.5&	25\fdg707,	0\fdg188&	10.895  $\pm$ 	0.065&	9.914  $\pm$ 	0.046	&6.644  $\pm$ 	0.043&	2.424	  $\pm$ 0.053&Class II, C3\\
29&J183731.09-060932.4&	25\fdg866,	0\fdg274&	10.868  $\pm$ 	0.049&	10.467  $\pm$ 	0.046&	7.521  $\pm$ 	0.065&	6.633  $\pm$ 	null&Class II, C1\\
30&J183727.40-061714.7	&25\fdg745,	0\fdg229&	10.352  $\pm$ 	0.067&	9.991	  $\pm$ 0.041&	5.598	  $\pm$ 0.041&	1.654  $\pm$ 	0.030&Class II, C2\\
31&J183728.07-061716.6&	25\fdg746,	0\fdg226&	10.597	  $\pm$ 0.062&	10.002  $\pm$ 	0.044	&5.443  $\pm$ 	0.058&	1.17	  $\pm$ 0.023&Class II, C2\\
32&J183726.23-061207.7	&25\fdg818,	0\fdg272&	10.883  $\pm$ 0.109	&10.26    $\pm$ 0.076&	5.504  $\pm$ 	0.033&	2.608  $\pm$ 	0.046&Class II, C1\\
33&J183724.66-061443.1	&25\fdg777,	0\fdg258&	11.037	  $\pm$ 0.076&	10.056	  $\pm$ 0.071&	5.243	  $\pm$ 0.032&	1.973	  $\pm$ 0.035&Class II, C1\\
34*&J183732.95-062018.7	&25\fdg710, 0.185&	8.882  $\pm$ 	0.028&	8.419	  $\pm$ 0.025&	4.519	  $\pm$ 0.021&	3.456  $\pm$ 	0.062&Class II, C3\\
35&J183725.03-061515.4	&25\fdg770,	0\fdg252	&10.402  $\pm$ 	0.037	&9.728  $\pm$ 	0.042&	5.351  $\pm$ 	0.053&	2.036  $\pm$ 	0.037&Class II, C1\\
36&J183730.25-062047.2&25\fdg698,	0\fdg191&	10.752  $\pm$ 	0.057&	10.298  $\pm$ 	0.049&	7.171  $\pm$ 	0.135&	1.568  $\pm$ 	0.026&Class II, C3\\
37&J183725.12-061500.3&	25\fdg774,	0\fdg254&	10.32	  $\pm$ 0.049&	9.947  $\pm$ 	0.065&	5.316  $\pm$ 	0.029&	2.108  $\pm$ 	0.040&Class II, C1\\
38&J183721.48-061309.7&25\fdg794,	0\fdg282&	9.442	  $\pm$ 0.031&	8.765	  $\pm$ 0.027&	6.646  $\pm$ 	0.032&	4.136	  $\pm$ 0.028&Class II, C1\\
39&J183723.96-061446.6&	25\fdg775,	0\fdg260	&11.475  $\pm$ 	0.109	&10.898	  $\pm$ 0.141&	7.26  $\pm$ 	0.148&	2.967  $\pm$ 	0.065&Class II, C1\\
40&J183805.93-060741.8	&25\fdg959, 0\fdg160	&10.944	$\pm$ 0.043	&10.569	$\pm$ 0.051	&7.426	$\pm$ 0.116&	4.443	$\pm$ 0.136&Class II, 100 \kms\\
41&J183810.23-061515.0 &	25\fdg856, 0\fdg087&	9.714 $\pm$	0.029&	9.305 $\pm$	0.031	&7.335 $\pm$ 0.086	&4.762 $\pm$ 0.117&Class II, 30 \kms\\
42&J183817.35-061815.5&	25\fdg825,	0\fdg0374	&9.912	$\pm$ 0.028&	9.525 $\pm$	0.031&	7.433 $\pm$	0.117	&4.574 $\pm$	0.060 &Class II, 100 \kms\\
43&J183747.97-061326.5	&25\fdg840, 0\fdg182&	9.771	$\pm$ 0.03&9.276	$\pm$ 0.026	&7.603	$\pm$ 0.134&	5.187	$\pm$ 0.136&Class II, 100 \kms\\
44&J183812.35-060918.5&	25\fdg948,	0\fdg124&8 .132 $\pm$  	0.035&	7.391 $\pm$ 	0.021&	6.202	$\pm$ 0.029	&4.151 $\pm$ 	0.039&Class II, 100 \kms\\
45&J183751.67-062025.4&	25\fdg744, 0\fdg115&	8.204$\pm$	0.026	&7.78$\pm$	0.023	&6.24 $\pm$	0.023&	1.543	$\pm$ 0.060&Class II, 30 and 100 \kms\\
46&J183745.91-060510.9&	25\fdg959, 0\fdg253&	8.994$\pm$	0.027&	8.518$\pm$	0.023	&7.088	$\pm$ 0.066	&4.981	$\pm$ 0.274&Class II, 100 \kms\\
47&J183812.05-061227.1	&25\fdg901,	0\fdg101&	10.487$\pm$	0.04&	9.721 $\pm$	0.036	&8.163 $\pm$	0.096	&6.295	$\pm$ 0.207&Class II, 100 \kms\\
48&J183810.70-060807.3	&25\fdg962,	0\fdg139&	10.579$\pm$ 	0.043&	10.255	$\pm$ 0.038	&7.502$\pm$ 	0.09&	5.793	$\pm$ 0.116&Class II, 100 \kms\\
49&J183815.45-061755.5&	25\fdg826,	0\fdg0470&	9.639$\pm$	0.079&	8.95$\pm$	0.07	&6.346$\pm$	0.118&	4.01$\pm$	0.081&Class II, 100 \kms\\
50&J183743.67-060833.7	&25\fdg905,	0\fdg235&	10.657$\pm$	0.037	&9.818$\pm$	0.037&	7.753$\pm$	0.115	&2.578$\pm$	0.027&Class II, 70 and  100 \kms\\
51& J183759.17-062006.6&	25\fdg763,	0\fdg090&	6.336	$\pm$	0.05	&5.796$\pm$		0.025&	3.14$\pm$		0.016&	1.373$\pm$		0.024&Class II,  30 and 100 \kms\\
52&J183757.75-061054.6	&25\fdg897,	0\fdg166&	6.126$\pm$		0.108&5.486$\pm$		0.06&	3.962$\pm$		0.017&	2.626	$\pm$	0.028&Class II, 65 and 100 \kms\\
53&J183752.61-061021.2&	25\fdg895,	0\fdg189	&10.814		$\pm$ 0.062&	10.115		$\pm$ 0.045	&8.106	$\pm$	0.102&	6.485		$\pm$ 0.144&Class II, 65 and 100 \kms\\
\hline
\end{tabular}
\end{table*}

\addtocounter{table}{-1}

\begin{table*}
\caption{ Continued.  \label{ysos}}
\begin{tabular}{c c c c c c c c}
\hline\hline
\multicolumn{8}{c}{{\bf Spitzer sources}} \\
$\#$ &  Designation &  {(\it l, b})& 4.5 $\mu$m [mag]  &   5.8 $\mu$m [mag]   &   8.0 $\mu$m [mag]  &    24 $\mu$m [mag]& Notes\\
\hline
54 &G025.7443+00.1845&	25\fdg744, 0\fdg184&	12.156 $\pm$	0.08&	11.027	$\pm$ 0.06&	10.16$\pm$ 0.072 &	9.955 $\pm$	0.146 & Class I\\
55&G025.7967+00.2764&	25\fdg796, 0\fdg276 &	12.3 $\pm$	0.058&	11.463	$\pm$ 0.098	&10.565 $\pm$	0.095&	9.796 $\pm$	0.113 & Class I, C1\\
56 &G025.7943+00.2827&	25\fdg794,	0\fdg282&9.937 $\pm$	0.05&9.146 $\pm$	0.043&	8.395	$\pm$ 0.039	&7.856	$\pm$ 0.045& Class I, C1\\
57&G025.7619+00.1029&	25\fdg762,	0\fdg103&	12.421 $\pm$	0.058&	10.408 $\pm$	0.11	&8.896 $\pm$	0.039&	8.025 $\pm$	0.046&Class I, 30, 60, and 100 \kms\\
58&G025.8895+00.0917	&25\fdg889, 0\fdg091&	13.512 $\pm$	0.088&	12.063 $\pm$	0.088&	11.088 $\pm$	0.108	&10.379 $\pm$	0.085&Class I, 100 and 30 \kms\\
59&G025.7978+00.0736&	25\fdg797,	0\fdg073&	12.572 $\pm$	0.08	&11.321 $\pm$	0.087&	10.416 $\pm$	0.074&	9.886 $\pm$	0.064&Class I, 30 and 60 \kms\\
60&G025.8564+00.0781&	25\fdg856,	0\fdg078	&12.506 $\pm$	0.096&	11.643 $\pm$	0.061	&10.878 $\pm$	0.094&	10.567 $\pm$	0.085&Class I, 100, 120, 60, and 30 \kms\\
61&G025.8320+00.0735	&25\fdg832, 0\fdg073&	13.609 $\pm$	0.128&	11.896 $\pm$	0.146&	10.567 $\pm$	0.07	&9.941 $\pm$	0.057&Class I, 30, 40, 100 \kms\\
62&G025.8991+00.1062&	25\fdg899,	0\fdg106&	12.735 $\pm$	0.079&	11.552 $\pm$	0.08&	10.507 $\pm$	0.083&	9.378 $\pm$	0.042&Class I, 100, 30, 60 \kms\\
63&G025.7545+00.2578	&25\fdg754,	0\fdg257&	11.246	0.077	&10.262	0.059&	9.564	0.055	&9.021	0.06&Class II, C2\\
64 &G025.7212+00.2476 &	25\fdg721,	0\fdg247&	8.775 $\pm$	0.043&	8.335 $\pm$	0.049&	7.761 $\pm$	0.033&	7.4	$\pm$ 0.032& Class II, C2\\
65&G025.8269+00.1224&	25\fdg827,	0\fdg122&	8.544 $\pm$	0.05&	8.289	$\pm$0.041	&7.76	0.034&	7.405	0.029& Class II, 90, 100, 30 \kms\\
66&G025.7900+00.1191&	25\fdg790,	0\fdg119&	12.4 $\pm$	0.078&	11.738	$\pm$0.074	&11.351$\pm$	0.1	&10.333	$\pm$0.079& Class II, 30 \kms\\
67&G025.7754+00.1288&	25\fdg775,	0\fdg128&	8.893 $\pm$	0.041&	8.602	$\pm$0.046	&8.338	$\pm$0.039	&7.925	$\pm$0.033& Class II, 30 \kms\\
68&G025.9598+00.1602&	25\fdg959,	0\fdg160&	10.815	$\pm$ 0.041&	10.412	$\pm$0.079&	10.134	$\pm$0.059	&9.238	$\pm$0.059& Class II, 100 \kms\\
69&G025.9303+00.1309&	25\fdg930,	0\fdg131&	10.868	$\pm$ 0.051&	10.298	$\pm$0.058&	9.749	$\pm$0.059&	8.926	$\pm$0.043& Class II, 100, 60 \kms\\
70&G025.7820+00.0632&	25\fdg782,	0\fdg063&	9.801	$\pm$ 0.06&	9.15	$\pm$0.051&	8.59	$\pm$0.039&	7.702$\pm$	0.029& Class II, 30, 110 \kms\\
71&G025.8833+00.2496&	25\fdg883,	0\fdg249&	10.020 $\pm$	0.045&	9.351	$\pm$0.053&	8.719	$\pm$0.036	&7.873$\pm$	0.048& Class II, 100, 70 \kms\\
72&G025.8383+00.2416&	25\fdg838,	0\fdg241&	12.622	$\pm$ 0.073&	12.022	$\pm$0.097&	11.685	$\pm$0.151&	10.532$\pm$	0.103& Class II, 100, 70 \kms\\
73&G025.8281+00.1596&	25\fdg828,	0\fdg159&	12.009 $\pm$	0.095&	11.106	$\pm$0.106	&10.454$\pm$	0.079&	9.781$\pm$	0.061& Class II, 100, 30, 60 \kms\\
74&G025.8786+00.1662&	25\fdg878,	0\fdg166&	11.467	$\pm$ 0.069&	11.035	$\pm$0.129&	10.685	$\pm$0.067&	10.099$\pm$	0.054& Class II, 60, 70, 100 \kms\\
75&G025.8865+00.1622&	25\fdg886,	0\fdg162&	11.923	$\pm$ 0.056&	11.406	$\pm$0.071&	10.775	$\pm$0.097&	10.2$\pm$	0.066& Class II,  60, 70, 100 \kms\\
76&G025.8133+00.1824&	25\fdg813,	0\fdg182	 &      11.854 $\pm$	0.111&	11.269	$\pm$0.079&	10.983	$\pm$0.073&	10.229$\pm$	0.08& Class II, 100, 60 \kms\\
77&G025.8548+00.1909&	25\fdg855,	0\fdg191&	12.254	$\pm$ 0.104&	11.604999	$\pm$0.116&	11.117	$\pm$0.088&	9.737$\pm$	0.081& Class II, 100, 60 \kms\\
78&G025.8952+00.1889&	25\fdg895,	0\fdg189&	10.405 $\pm$	0.04&	9.823	$\pm$0.057&	9.148	$\pm$0.044	&8.433	$\pm$0.034& Class II, 100, 60, 70 \kms\\
\hline
\end{tabular}
\end{table*}

\subsubsection{Spatial distribution of cYSOs and possible formation scenarios}

Figure \ref{fig-ysos}  shows that  more than half of the  cYSOS are  found   located  onto the PDR,  with enhanced  concentration  at  the  locations  of  the molecular clumps C1, C2, C3 and C4  (see  Column 8 of Table \ref{ysos}), while the rest are seen located projected toward the inner part of the \hii\, region where only faint molecular gas emission is detected at the velocity interval between 100 and 115 \kms\,, as delineated by the 0.3 K contour in Fig. \ref{fig-ysos}.

On the other hand, we note that in the direction of the cYSOs located onto C1, C2, C3, and C4  most of the CO profiles show one single radial component and that in the cases where two or more components are detected, their intensities are considerably lower than the component within the velocity interval between +108 and +120 \kms.  On the contrary, in direction to the rest of the cYSOs, several CO components are observed, as indicated in the last column of Table \ref{ysos}, where the radial velocities of the observed components are shown in decreasing order according to their relative intensities. This suggests that several cYSOs are probably related to  molecular gas located at a different distance than \g, although the possibility that either  they have already destroyed all their natal cloud  or that they are not protostellar objects cannot be discarded. 

The observed spatial distribution of the cYSOs is a good indicator that the action of \g\, onto its environs is strong and could have triggered the formation of new stars. The estimated column densities (see Table \ref{coprop}  and \ref{mass-bolocam}) are all above the threshold found by \citet{and11}, $N_{H2} \geq 7 \times 10^{21}$\, cm$^{-2}$, indicating that star formation may well have taken place in all four clumps. Moreover, the mass-size relationship, which establishes that if $m(r) \geq 870 M_{\odot}\,(r/pc)^{1.33}$ the cloud may form high-mass stars \citep{kau10}, is fulfilled in two of them, C1 and C2.

\paragraph{The Radiative driven implosion scenario.}
An close inspection of Fig. \ref{fig-ysos} suggests that 
 the radiation-driven implosion (RDI), could be at work in this region, since the molecular clump labeled C1 exhibits a cometary shape (see Sect. \ref{bolocam}).

Radiative pressure from \hii\, regions push away low density gas more efficiently, and hence faster, than high density gas, thus perturbing the structure and dynamics of molecular clouds. As a result, when an \hii\, region finds dense molecular clumps along its expansion, it may give rise to the formation of bright rim clouds. 
When this happens,  an IBL is  developed around the clump and a photo-ionization front is driven into it. The subsequent evolution of the cloud is determined by the pressure balance between the internal molecular pressure and the external pressure of the IBL \citep{lef94}.

\begin{figure}
\centering
\includegraphics[width=9cm]{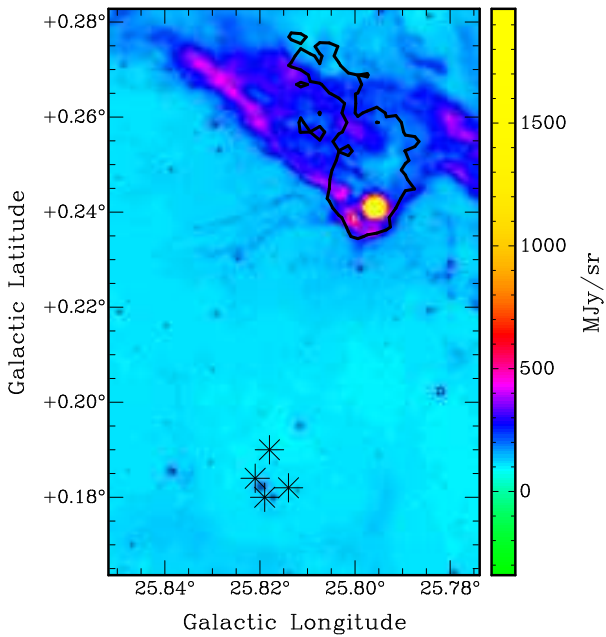}
\caption{Spitzer image at 8 $\mu$m of the region of the C1 source. The contour indicates the 1.1 mm emission at 0.3 Jy\, beam$^{-1}$. The four asterisks show the location of the 2MASS sources candidates of being O-type stars (\#16, 17, 18, and 19 in Table \ref{estrellas} and Fig. \ref{reg}).
  }
\label{rdi}
\end{figure}

Figure \ref{rdi} shows an enlarged view of the region of the clump C1.
The cometary shape of the molecular cloud, shown through its BOLOCAM emission, is evident, as well as its bright infrared border. 
 Given their location with respect to the head of C1, the 2MASS sources \#16, 17, 18 and 19 (asterisks in Fig. \ref{rdi}) are candidates of being the O-type stars creating the IBL.

The cometary shape of the  structure suggests that the RDI mechanism could be at work  in this molecular cloud.
Moreover, the presence of several cYSOs projected onto C1 (see Fig. \ref{fig-ysos}) suggests that a star cluster has formed there and one of them has already given rise to the CHII region G025.7961+00.2403. To ascertain whether this is the case, both inner and outer pressures acting onto the clump have to be estimated and compared.

The pressure in the IBL can be evaluated from the electron density in the boundary layer, $n_e$, as $P_i = 2\, n_e \, m_H\, c_i^2$, where  $m_H = 1.67 \times 10^{-24}$g, is  the mass of the hydrogen atom and $c_i$ is the sound speed in the ionized gas ($c_i \sim$ 14 \kms). A rough estimation of $n_e$ can be obtained using the spherical model of \citet{mez67} with an appropriate filling factor $f$ to account for the fraction of the assumed sphere that  actually contains ionized gas,

$$n_e  =  533.5 \,T_4^{0.175}\,\nu_{\rm GHz}
           ^{0.05}\,S_{\nu}^{0.5}\,(f^{1/3}\,\theta)^{-1.5}\,D_{\rm kpc}
           ^{-0.5} \rm\, cm^{-3},$$
           
\noindent where $S_{\nu}$ is the measured flux density in Jy, $\theta$ 
the angular width in minutes of arc, $T_4$ the electron temperature 
in units of $10^4$ K, $D_{\rm kpc}$ the distance in kpc, and $\nu_{\rm GHz}$ the frequency in GHz.

Based on the MAGPIS 1.42 GHz image, we obtained a flux density $S_{1.42} = 120 \pm 30$ mJy for the emission observed in the border of C1.
Then, assuming a distance of 6.5 $\pm 1.0$ kpc,  a 
temperature of $6120 \pm 100$ K, the C1 angular width, $\theta = 4 \pm 1 $ arcmin and $f=0.03$, we obtain $n_e= 70 \pm 30$ cm$^{-3}$ and $P_i/k = (3.3 \pm 1.4) \times 10^6$ cm$^{-3}\,K$.

As mentioned above, this pressure should be compared with the pressure of the molecular cloud, which can be estimated, assuming that the thermal component can be neglected, from the turbulent velocity dispersion, $\sigma^2$, and the molecular density, $\rho_m$, as $P_m = \sigma^2 \, \rho_m$, where  $\sigma^2$ may be written as $\sigma^2 = \Delta\, v^2/$(8\,ln 2), $\Delta\,v$ being the observed velocity line width of the molecular cloud gas.
Adopting for C1 the ambient density given in Table \ref{coprop}, $n_{H2} = 2.1 \times 10^3$ cm$^{-3}$, and  $\Delta\,v = 4.8 \pm 0.7$ \kms, we infer a molecular pressure $P_m/k = (2.8 \pm 1.2) \times 10^{-6}$ cm$^{-3}\,$K, assuming a 30\% error in the molecular density.

Since the pressures $P_i$ and  $P_m$ obtained for C1 are similar,  the ionized and molecular gas seem to be in pressure balance and the propagation of a photoionization-induced shock into the molecular gas  might have occurred.

\section{Summary}

We have carried out a multi-wavelength study of the ionized region \g\, in order to characterize the physical properties of the related gas and dust and to investigate its possible origin and role in forming new stars. Based on data obtained from several surveys we arrived to the following results:

\begin{enumerate}

\item Since \g\, is located in the first Galactic quadrant and close to the tangent point, its distance was not clear.
Based on the comparison of \hi\, absorption profiles, in this work we could solve the distance ambiguity and determine a distance of 6.5 $\pm 1.0$ kpc from the Sun, which corresponds to the near distance.

\item The region is very bright at 1420 MHz radio continuum. It contains about 2500 M$_{\odot}$ of ionized gas, and at least $5 \times 10^{49}$ ph\,s$^{-1}$ are needed to keep it ionized. This implies that several O-type stars, which have not been detected yet, should be located in this part of the Galaxy.  

\item The $^{13}$CO\,(1-0) and 1.1\,mm data reveal the presence of four clumps bordering the \hii\, region. These are detected approximately in the velocity range from 106 to 116 \kms, compatible with the velocity of the RRLs detected in the region. Based on  their estimated physical properties, we conclude that all these clumps are capable of forming new stars and, in particular, clumps C1 and C2 may contain high-mass protostars. 

\item At infrared wavelengths, we identified an unmistakable photo-dissociation region in the interface between the ionized and molecular gas. The Herschel data show that the region is also detected in the far infrared. We found that the dust temperature goes from about 21 K in the cold clumps, up to 31 K in the region more directly exposed to stellar radiation.

\item Since \g\, is located in the inner galaxy, where the visual extinction is high, the candidate ionizing stars had to be searched in the infrared. 
There is one WR star, HDM\,40, and one RSG, 18375756-0620155, that may be related to the region. On the other hand, based on color criteria, we identified 37 2MASS sources which could be the O-type ionizing stars.  

\item Applying  different colour criteria,  78 YSO candidates have been identified in the region, of which 44 are projected  onto the molecular gas related to \g.
Based on the observed distribution of both the infrared and molecular emission, we propose that the RDI mechanism has taken place onto C1 and triggered the formation of the high-mass star responsible of the  striking CHII region G025.7961+00.2403.

\end{enumerate}

\section*{Acknowledgments}
This work is dedicated to the memory of our dear colleague Dr. Gisela Romero.
{\bf We are grateful to the referee for his/her very constructive input.}
The VGPS is supported by a grant from the Natural Sciences             
and Engineering Research Council of Canada and from the U.S.            
 National Science Foundation. The National Radio                         
 Astronomy Observatory is a facility of the National Science             
 Foundation operated under cooperative agreement by Associated           
 Universities,Inc.
 This research has made use of the NASA/ IPAC Infrared Science Archive, which is operated by the Jet Propulsion Laboratory, California Institute of Technology, under contract with the National Aeronautics and Space Administration. Herschel is an ESA space observatory with science instruments provided by European-led Principal Investigator consortia and with important participation from NASA. This work is partially based on observations made with the Spitzer Space Telescope, which is operated by the Jet Propulsion Laboratory, California Institute of Technology under a contract with NASA. This publication makes use of molecular line data from the Boston University-FCRAO Galactic Ring Survey (GRS). The GRS is a joint project of Boston University and Five College Radio Astronomy Observatory, funded by the National Science Foundation under grants AST-9800334, AST-0098562, $\&$ AST-0100793 
 This project was partially financed by the Consejo Nacional de Investigaciones
Cientificas y Tecnicas (CONICET) of Argentina under project PIP\,01299, PIP\,0226, PIP\,00356, and  PIP\,00107, Universidad de La Plata UNLP 2012-2014 PPID/G002 and 11/G120. 

\bibliographystyle{mn2e}  
\bibliography{bib-todo}

\begin{thebibliography}{66}
\expandafter\ifx\csname natexlab\endcsname\relax\def\natexlab#1{#1}\fi

\bibitem[{{Aguirre} {et~al}\mbox{.}(2011){Aguirre}, {Ginsburg}, {Dunham},
  {Drosback}, {Bally}, {Battersby}, {Bradley}, {Cyganowski}, {Dowell}, {Evans},
  {Glenn}, {Harvey}, {Rosolowsky}, {Stringfellow}, {Walawender}, \&
  {Williams}}]{agui11}
{Aguirre} J.~E. {et~al.}, 2011, \apjs, 192, 4

\bibitem[{{Anderson} \& {Bania}(2009)}]{anda09}
{Anderson} L.~D., {Bania} T.~M., 2009, \apj, 690, 706

\bibitem[{{Anderson} {et~al}\mbox{.}(2014){Anderson}, {Bania}, {Balser},
  {Cunningham}, {Wenger}, {Johnstone}, \& {Armentrout}}]{and14}
{Anderson} L.~D., {Bania} T.~M., {Balser} D.~S., {Cunningham} V., {Wenger}
  T.~V., {Johnstone} B.~M., {Armentrout} W.~P., 2014, \apjs, 212, 1

\bibitem[{{Anderson} {et~al}\mbox{.}(2009){Anderson}, {Bania}, {Jackson},
  {Clemens}, {Heyer}, {Simon}, {Shah}, \& {Rathborne}}]{andb09}
{Anderson} L.~D., {Bania} T.~M., {Jackson} J.~M., {Clemens} D.~P., {Heyer} M.,
  {Simon} R., {Shah} R.~Y., {Rathborne} J.~M., 2009, \apjs, 181, 255

\bibitem[{{Andr{\'e}} {et~al}\mbox{.}(2011){Andr{\'e}}, {Men'shchikov},
  {K{\"o}nyves}, \& {Arzoumanian}}]{and11}
{Andr{\'e}} P., {Men'shchikov} A., {K{\"o}nyves} V., {Arzoumanian} D., 2011, in
  IAU Symposium, Vol. 270, Computational Star Formation, {Alves} J.,
  {Elmegreen} B.~G., {Girart} J.~M., {Trimble} V., eds., pp. 255--262

\bibitem[{{Benjamin} {et~al}\mbox{.}(2003){Benjamin}, {Churchwell}, {Babler},
  {Bania}, {Clemens}, {Cohen}, {Dickey}, {Indebetouw}, {Jackson}, {Kobulnicky},
  {Lazarian}, {Marston}, {Mathis}, {Meade}, {Seager}, {Stolovy}, {Watson},
  {Whitney}, {Wolff}, \& {Wolfire}}]{ben03}
{Benjamin} R.~A. {et~al.}, 2003, \pasp, 115, 953

\bibitem[{{Cazzolato} \& {Pineault}(2005)}]{caz05}
{Cazzolato} F., {Pineault} S., 2005, \aj, 129, 2731

\bibitem[{{Chaisson}(1976)}]{cha76}
{Chaisson} E.~J., 1976, in Frontiers of Astrophysics, {Avrett} E.~H., ed., pp.
  259--351

\bibitem[{{Chen} {et~al}\mbox{.}(1999){Chen}, {Figueras}, {Torra}, {Jordi},
  {Luri}, \& {Galad{\'{\i}}-Enr{\'{\i}}quez}}]{che99}
{Chen} B., {Figueras} F., {Torra} J., {Jordi} C., {Luri} X.,
  {Galad{\'{\i}}-Enr{\'{\i}}quez} D., 1999, \aap, 352, 459

\bibitem[{{Comer{\'o}n} {et~al}\mbox{.}(2002){Comer{\'o}n}, {Pasquali},
  {Rodighiero}, {Stanishev}, {De Filippis}, {L{\'o}pez Mart{\'{\i}}},
  {G{\'a}lvez Ortiz}, {Stankov}, \& {Gredel}}]{com02}
{Comer{\'o}n} F. {et~al.}, 2002, \aap, 389, 874

\bibitem[{{Cutri} {et~al}\mbox{.}(2003){Cutri}, {Skrutskie}, {van Dyk},
  {Beichman}, {Carpenter}, {Chester}, {Cambresy}, {Evans}, {Fowler}, {Gizis},
  {Howard}, {Huchra}, {Jarrett}, {Kopan}, {Kirkpatrick}, {Light}, {Marsh},
  {McCallon}, {Schneider}, {Stiening}, {Sykes}, {Weinberg}, {Wheaton},
  {Wheelock}, \& {Zacarias}}]{cut03}
{Cutri} R.~M. {et~al.}, 2003, {2MASS All Sky Catalog of point sources.} The
  IRSA 2MASS All-Sky Point Source Catalog, NASA/IPAC Infrared Science
  Archive.~http://irsa.ipac.caltech.edu/applications/Gator/

\bibitem[{{Dame}, {Hartmann} \& {Thaddeus}(2001){Dame}, {Hartmann}, \&
  {Thaddeus}}]{dame01}
{Dame} T.~M., {Hartmann} D., {Thaddeus} P., 2001, \apj, 547, 792

\bibitem[{{Davies} {et~al}\mbox{.}(2007){Davies}, {Figer}, {Kudritzki},
  {MacKenty}, {Najarro}, \& {Herrero}}]{dav07}
{Davies} B., {Figer} D.~F., {Kudritzki} R.-P., {MacKenty} J., {Najarro} F.,
  {Herrero} A., 2007, \apj, 671, 781

\bibitem[{{Davies} {et~al}\mbox{.}(2008){Davies}, {Figer}, {Law}, {Kudritzki},
  {Najarro}, {Herrero}, \& {MacKenty}}]{dav08}
{Davies} B., {Figer} D.~F., {Law} C.~J., {Kudritzki} R.-P., {Najarro} F.,
  {Herrero} A., {MacKenty} J.~W., 2008, \apj, 676, 1016

\bibitem[{{Dempsey}, {Thomas} \& {Currie}(2013){Dempsey}, {Thomas}, \&
  {Currie}}]{demp13}
{Dempsey} J.~T., {Thomas} H.~S., {Currie} M.~J., 2013, \apjs, 209, 8

\bibitem[{{Dickman}(1978)}]{d78}
{Dickman} R.~L., 1978, \apjs, 37, 407

\bibitem[{{Dorda} {et~al}\mbox{.}(2016){Dorda}, {Negueruela},
  {Gonz{\'a}lez-Fern{\'a}ndez}, \& {Marco}}]{dor16}
{Dorda} R., {Negueruela} I., {Gonz{\'a}lez-Fern{\'a}ndez} C., {Marco} A., 2016,
  in Astronomical Society of the Pacific Conference Series, Vol. 507,
  Multi-Object Spectroscopy in the Next Decade: Big Questions, Large Surveys,
  and Wide Fields, {Skillen} I., {Barcells} M., {Trager} S., eds., p. 165

\bibitem[{{Draine} {et~al}\mbox{.}(2007){Draine}, {Dale}, {Bendo}, {Gordon},
  {Smith}, {Armus}, {Engelbracht}, {Helou}, {Kennicutt}, {Li}, {Roussel},
  {Walter}, {Calzetti}, {Moustakas}, {Murphy}, {Rieke}, {Bot}, {Hollenbach},
  {Sheth}, \& {Teplitz}}]{dra07}
{Draine} B.~T. {et~al.}, 2007, \apj, 663, 866

\bibitem[{{Drilling} \& {Landolt}(2000)}]{dri00}
{Drilling} J.~S., {Landolt} A.~U., 2000, {Normal Stars}, {Cox} A.~N., ed., p.
  381

\bibitem[{{Egan}, {Price} \& {Kraemer}(2003){Egan}, {Price}, \&
  {Kraemer}}]{ega03}
{Egan} M.~P., {Price} S.~D., {Kraemer} K.~E., 2003, in Bulletin of the American
  Astronomical Society, Vol.~35, American Astronomical Society Meeting
  Abstracts, p. 1301

\bibitem[{{Ekstr{\"o}m} {et~al}\mbox{.}(2012){Ekstr{\"o}m}, {Georgy},
  {Eggenberger}, {Meynet}, {Mowlavi}, {Wyttenbach}, {Granada}, {Decressin},
  {Hirschi}, {Frischknecht}, {Charbonnel}, \& {Maeder}}]{eks2012}
{Ekstr{\"o}m} S. {et~al.}, 2012, \aap, 537, A146

\bibitem[{{Ekstr{\"o}m} {et~al}\mbox{.}(2013){Ekstr{\"o}m}, {Georgy}, {Meynet},
  {Groh}, \& {Granada}}]{eks2013}
{Ekstr{\"o}m} S., {Georgy} C., {Meynet} G., {Groh} J., {Granada} A., 2013, in
  EAS Publications Series, Vol.~60, EAS Publications Series, {Kervella} P., {Le
  Bertre} T., {Perrin} G., eds., pp. 31--41

\bibitem[{{Elmegreen} \& {Lada}(1977)}]{elm77}
{Elmegreen} B.~G., {Lada} C.~J., 1977, \apj, 214, 725

\bibitem[{{Fich}, {Blitz} \& {Stark}(1989){Fich}, {Blitz}, \& {Stark}}]{fbs89}
{Fich} M., {Blitz} L., {Stark} A.~A., 1989, \apj, 342, 272

\bibitem[{{Figer} {et~al}\mbox{.}(2006){Figer}, {MacKenty}, {Robberto},
  {Smith}, {Najarro}, {Kudritzki}, \& {Herrero}}]{fig06}
{Figer} D.~F., {MacKenty} J.~W., {Robberto} M., {Smith} K., {Najarro} F.,
  {Kudritzki} R.~P., {Herrero} A., 2006, \apj, 643, 1166

\bibitem[{{Glenn} {et~al}\mbox{.}(2003){Glenn}, {Ade}, {Amarie}, {Bock},
  {Edgington}, {Goldin}, {Golwala}, {Haig}, {Lange}, {Laurent}, {Mauskopf},
  {Yun}, \& {Nguyen}}]{glen03}
{Glenn} J. {et~al.}, 2003, in Millimeter and Submillimeter Detectors for
  Astronomy, {Phillips} T.~G., {Zmuidzinas} J., eds., Vol. 4855, pp. 30--40

\bibitem[{{Griffin} {et~al}\mbox{.}(2010){Griffin}, {Abergel}, {Abreu}, {Ade},
  {Andr{\'e}}, {Augueres}, {Babbedge}, {Bae}, {Baillie}, {Baluteau}, {Barlow},
  {Bendo}, {Benielli}, {Bock}, {Bonhomme}, {Brisbin}, {Brockley-Blatt},
  {Caldwell}, {Cara}, {Castro-Rodriguez}, {Cerulli}, {Chanial}, {Chen},
  {Clark}, {Clements}, {Clerc}, {Coker}, {Communal}, {Conversi}, {Cox},
  {Crumb}, {Cunningham}, {Daly}, {Davis}, {de Antoni}, {Delderfield}, {Devin},
  {di Giorgio}, {Didschuns}, {Dohlen}, {Donati}, {Dowell}, {Dowell}, {Duband},
  {Dumaye}, {Emery}, {Ferlet}, {Ferrand}, {Fontignie}, {Fox}, {Franceschini},
  {Frerking}, {Fulton}, {Garcia}, {Gastaud}, {Gear}, {Glenn}, {Goizel},
  {Griffin}, {Grundy}, {Guest}, {Guillemet}, {Hargrave}, {Harwit}, {Hastings},
  {Hatziminaoglou}, {Herman}, {Hinde}, {Hristov}, {Huang}, {Imhof}, {Isaak},
  {Israelsson}, {Ivison}, {Jennings}, {Kiernan}, {King}, {Lange}, {Latter},
  {Laurent}, {Laurent}, {Leeks}, {Lellouch}, {Levenson}, {Li}, {Li},
  {Lilienthal}, {Lim}, {Liu}, {Lu}, {Madden}, {Mainetti}, {Marliani}, {McKay},
  {Mercier}, {Molinari}, {Morris}, {Moseley}, {Mulder}, {Mur}, {Naylor},
  {Nguyen}, {O'Halloran}, {Oliver}, {Olofsson}, {Olofsson}, {Orfei}, {Page},
  {Pain}, {Panuzzo}, {Papageorgiou}, {Parks}, {Parr-Burman}, {Pearce},
  {Pearson}, {P{\'e}rez-Fournon}, {Pinsard}, {Pisano}, {Podosek}, {Pohlen},
  {Polehampton}, {Pouliquen}, {Rigopoulou}, {Rizzo}, {Roseboom}, {Roussel},
  {Rowan-Robinson}, {Rownd}, {Saraceno}, {Sauvage}, {Savage}, {Savini},
  {Sawyer}, {Scharmberg}, {Schmitt}, {Schneider}, {Schulz}, {Schwartz},
  {Shafer}, {Shupe}, {Sibthorpe}, {Sidher}, {Smith}, {Smith}, {Smith},
  {Spencer}, {Stobie}, {Sudiwala}, {Sukhatme}, {Surace}, {Stevens}, {Swinyard},
  {Trichas}, {Tourette}, {Triou}, {Tseng}, {Tucker}, {Turner}, {Vaccari},
  {Valtchanov}, {Vigroux}, {Virique}, {Voellmer}, {Walker}, {Ward}, {Waskett},
  {Weilert}, {Wesson}, {White}, {Whitehouse}, {Wilson}, {Winter}, {Woodcraft},
  {Wright}, {Xu}, {Zavagno}, {Zemcov}, {Zhang}, \& {Zonca}}]{gri10}
{Griffin} M.~J. {et~al.}, 2010, \aap, 518, L3

\bibitem[{{Gutermuth} {et~al}\mbox{.}(2009){Gutermuth}, {Megeath}, {Myers},
  {Allen}, {Pipher}, \& {Fazio}}]{gut09}
{Gutermuth} R.~A., {Megeath} S.~T., {Myers} P.~C., {Allen} L.~E., {Pipher}
  J.~L., {Fazio} G.~G., 2009, \apjs, 184, 18

\bibitem[{{Helfand} {et~al}\mbox{.}(2006){Helfand}, {Becker}, {White},
  {Fallon}, \& {Tuttle}}]{hel06}
{Helfand} D.~J., {Becker} R.~H., {White} R.~L., {Fallon} A., {Tuttle} S., 2006,
  \aj, 131, 2525

\bibitem[{{Higuchi} {et~al}\mbox{.}(2013){Higuchi}, {Kurono}, {Naoi}, {Saito},
  {Mauersberger}, \& {Kawabe}}]{hig13}
{Higuchi} A.~E., {Kurono} Y., {Naoi} T., {Saito} M., {Mauersberger} R.,
  {Kawabe} R., 2013, \apj, 765, 101

\bibitem[{{Hildebrand}(1983)}]{hil83}
{Hildebrand} R.~H., 1983, Quarterly Journal of the Royal Astronomical Society,
  24, 267

\bibitem[{{Hollenbach} \& {Tielens}(1997)}]{ht97}
{Hollenbach} D.~J., {Tielens} A.~G.~G.~M., 1997, \araa, 35, 179

\bibitem[{{Jackson} {et~al}\mbox{.}(2006){Jackson}, {Rathborne}, {Shah},
  {Simon}, {Bania}, {Clemens}, {Chambers}, {Johnson}, {Dormody}, {Lavoie}, \&
  {Heyer}}]{jac06}
{Jackson} J.~M. {et~al.}, 2006, \apjs, 163, 145

\bibitem[{{Kauffmann} \& {Pillai}(2010)}]{kau10}
{Kauffmann} J., {Pillai} T., 2010, \apjl, 723, L7

\bibitem[{{Koenig} {et~al}\mbox{.}(2012){Koenig}, {Leisawitz}, {Benford},
  {Rebull}, {Padgett}, \& {Assef}}]{koe12}
{Koenig} X.~P., {Leisawitz} D.~T., {Benford} D.~J., {Rebull} L.~M., {Padgett}
  D.~L., {Assef} R.~J., 2012, \apj, 744, 130

\bibitem[{{Kos} {et~al}\mbox{.}(2014){Kos}, {Zwitter}, {Wyse}, {Bienaym{\'e}},
  {Binney}, {Bland-Hawthorn}, {Freeman}, {Gibson}, {Gilmore}, {Grebel},
  {Helmi}, {Kordopatis}, {Munari}, {Navarro}, {Parker}, {Reid}, {Seabroke},
  {Sharma}, {Siebert}, {Siviero}, {Steinmetz}, {Watson}, \& {Williams}}]{kos14}
{Kos} J. {et~al.}, 2014, Science, 345, 791

\bibitem[{{Lefloch} \& {Lazareff}(1994)}]{lef94}
{Lefloch} B., {Lazareff} B., 1994, \aap, 289, 559

\bibitem[{{Levesque} {et~al}\mbox{.}(2005){Levesque}, {Massey}, {Olsen},
  {Plez}, {Josselin}, {Maeder}, \& {Meynet}}]{lev05}
{Levesque} E.~M., {Massey} P., {Olsen} K.~A.~G., {Plez} B., {Josselin} E.,
  {Maeder} A., {Meynet} G., 2005, \apj, 628, 973

\bibitem[{{Lockman}, {Pisano} \& {Howard}(1996){Lockman}, {Pisano}, \&
  {Howard}}]{loc96}
{Lockman} F.~J., {Pisano} D.~J., {Howard} G.~J., 1996, \apj, 472, 173

\bibitem[{{Lumsden} {et~al}\mbox{.}(2002){Lumsden}, {Hoare}, {Oudmaijer}, \&
  {Richards}}]{lum02}
{Lumsden} S.~L., {Hoare} M.~G., {Oudmaijer} R.~D., {Richards} D., 2002, \mnras,
  336, 621

\bibitem[{{Martins} \& {Plez}(2006)}]{mar06}
{Martins} F., {Plez} B., 2006, \aap, 457, 637

\bibitem[{{Martins}, {Schaerer} \& {Hillier}(2005){Martins}, {Schaerer}, \&
  {Hillier}}]{mar05}
{Martins} F., {Schaerer} D., {Hillier} D.~J., 2005, \aap, 436, 1049

\bibitem[{{Mauerhan}, {Van Dyk} \& {Morris}(2011){Mauerhan}, {Van Dyk}, \&
  {Morris}}]{mau11}
{Mauerhan} J.~C., {Van Dyk} S.~D., {Morris} P.~W., 2011, \aj, 142, 40

\bibitem[{{Mezger} \& {Henderson}(1967)}]{mez67}
{Mezger} P.~G., {Henderson} A.~P., 1967, \apj, 147, 471

\bibitem[{{Molinari} {et~al}\mbox{.}(2010){Molinari}, {Swinyard}, {Bally},
  {Barlow}, {Bernard}, {Martin}, {Moore}, {Noriega-Crespo}, {Plume}, {Testi},
  {Zavagno}, {Abergel}, {Ali}, {Andr{\'e}}, {Baluteau}, {Benedettini},
  {Bern{\'e}}, {Billot}, {Blommaert}, {Bontemps}, {Boulanger}, {Brand},
  {Brunt}, {Burton}, {Campeggio}, {Carey}, {Caselli}, {Cesaroni}, {Cernicharo},
  {Chakrabarti}, {Chrysostomou}, {Codella}, {Cohen}, {Compiegne}, {Davis}, {de
  Bernardis}, {de Gasperis}, {Di Francesco}, {di Giorgio}, {Elia}, {Faustini},
  {Fischera}, {Fukui}, {Fuller}, {Ganga}, {Garcia-Lario}, {Giard}, {Giardino},
  {Glenn}, {Goldsmith}, {Griffin}, {Hoare}, {Huang}, {Jiang}, {Joblin},
  {Joncas}, {Juvela}, {Kirk}, {Lagache}, {Li}, {Lim}, {Lord}, {Lucas},
  {Maiolo}, {Marengo}, {Marshall}, {Masi}, {Massi}, {Matsuura}, {Meny},
  {Minier}, {Miville-Desch{\^e}nes}, {Montier}, {Motte}, {M{\"u}ller},
  {Natoli}, {Neves}, {Olmi}, {Paladini}, {Paradis}, {Pestalozzi}, {Pezzuto},
  {Piacentini}, {Pomar{\`e}s}, {Popescu}, {Reach}, {Richer}, {Ristorcelli},
  {Roy}, {Royer}, {Russeil}, {Saraceno}, {Sauvage}, {Schilke},
  {Schneider-Bontemps}, {Schuller}, {Schultz}, {Shepherd}, {Sibthorpe},
  {Smith}, {Smith}, {Spinoglio}, {Stamatellos}, {Strafella}, {Stringfellow},
  {Sturm}, {Taylor}, {Thompson}, {Tuffs}, {Umana}, {Valenziano}, {Vavrek},
  {Viti}, {Waelkens}, {Ward-Thompson}, {White}, {Wyrowski}, {Yorke}, \&
  {Zhang}}]{mol10}
{Molinari} S. {et~al.}, 2010, \pasp, 122, 314

\bibitem[{{Negueruela} {et~al}\mbox{.}(2012){Negueruela}, {Marco},
  {Gonz{\'a}lez-Fern{\'a}ndez}, {Jim{\'e}nez-Esteban}, {Clark}, {Garcia}, \&
  {Solano}}]{neg12}
{Negueruela} I., {Marco} A., {Gonz{\'a}lez-Fern{\'a}ndez} C.,
  {Jim{\'e}nez-Esteban} F., {Clark} J.~S., {Garcia} M., {Solano} E., 2012,
  \aap, 547, A15

\bibitem[{{Oka} {et~al}\mbox{.}(2012){Oka}, {Onodera}, {Nagai}, {Tanaka},
  {Matsumura}, \& {Kamegai}}]{oka12}
{Oka} T., {Onodera} Y., {Nagai} M., {Tanaka} K., {Matsumura} S., {Kamegai} K.,
  2012, \apjs, 201, 14

\bibitem[{{Ossenkopf} \& {Henning}(1994)}]{osse94}
{Ossenkopf} V., {Henning} T., 1994, \aap, 291, 943

\bibitem[{{Poglitsch} {et~al}\mbox{.}(2010){Poglitsch}, {Waelkens}, {Geis},
  {Feuchtgruber}, {Vandenbussche}, {Rodriguez}, {Krause}, {Renotte}, {van
  Hoof}, {Saraceno}, {Cepa}, {Kerschbaum}, {Agn{\`e}se}, {Ali}, {Altieri},
  {Andreani}, {Augueres}, {Balog}, {Barl}, {Bauer}, {Belbachir}, {Benedettini},
  {Billot}, {Boulade}, {Bischof}, {Blommaert}, {Callut}, {Cara}, {Cerulli},
  {Cesarsky}, {Contursi}, {Creten}, {De Meester}, {Doublier}, {Doumayrou},
  {Duband}, {Exter}, {Genzel}, {Gillis}, {Gr{\"o}zinger}, {Henning},
  {Herreros}, {Huygen}, {Inguscio}, {Jakob}, {Jamar}, {Jean}, {de Jong},
  {Katterloher}, {Kiss}, {Klaas}, {Lemke}, {Lutz}, {Madden}, {Marquet},
  {Martignac}, {Mazy}, {Merken}, {Montfort}, {Morbidelli}, {M{\"u}ller},
  {Nielbock}, {Okumura}, {Orfei}, {Ottensamer}, {Pezzuto}, {Popesso},
  {Putzeys}, {Regibo}, {Reveret}, {Royer}, {Sauvage}, {Schreiber}, {Stegmaier},
  {Schmitt}, {Schubert}, {Sturm}, {Thiel}, {Tofani}, {Vavrek}, {Wetzstein},
  {Wieprecht}, \& {Wiezorrek}}]{pog10}
{Poglitsch} A. {et~al.}, 2010, \aap, 518, L2

\bibitem[{{Quireza} {et~al}\mbox{.}(2006{\natexlab{a}}){Quireza}, {Rood},
  {Balser}, \& {Bania}}]{quia06}
{Quireza} C., {Rood} R.~T., {Balser} D.~S., {Bania} T.~M., 2006{\natexlab{a}},
  \apjs, 165, 338

\bibitem[{{Quireza} {et~al}\mbox{.}(2006{\natexlab{b}}){Quireza}, {Rood},
  {Bania}, {Balser}, \& {Maciel}}]{quib06}
{Quireza} C., {Rood} R.~T., {Bania} T.~M., {Balser} D.~S., {Maciel} W.~J.,
  2006{\natexlab{b}}, \apj, 653, 1226

\bibitem[{{Reynoso}, {Cichowolski} \& {Walsh}(2017){Reynoso}, {Cichowolski}, \&
  {Walsh}}]{rcw17}
{Reynoso} E.~M., {Cichowolski} S., {Walsh} A.~J., 2017, \mnras, 464, 3029

\bibitem[{{Rieke} \& {Lebofsky}(1985)}]{rie85}
{Rieke} G.~H., {Lebofsky} M.~J., 1985, \apj, 288, 618

\bibitem[{{Rieke} {et~al}\mbox{.}(2004){Rieke}, {Young}, {Engelbracht},
  {Kelly}, {Low}, {Haller}, {Beeman}, {Gordon}, {Stansberry}, \&
  {Misselt}}]{rie04}
{Rieke} G.~H. {et~al.}, 2004, \apjs, 154, 25

\bibitem[{{Schlafly} \& {Finkbeiner}(2011)}]{sch11}
{Schlafly} E.~F., {Finkbeiner} D.~P., 2011, \apj, 737, 103

\bibitem[{{Schnee} {et~al}\mbox{.}(2005){Schnee}, {Ridge}, {Goodman}, \&
  {Li}}]{sch05}
{Schnee} S.~L., {Ridge} N.~A., {Goodman} A.~A., {Li} J.~G., 2005, \apj, 634,
  442

\bibitem[{{Sewilo} {et~al}\mbox{.}(2004){Sewilo}, {Watson}, {Araya},
  {Churchwell}, {Hofner}, \& {Kurtz}}]{sew04}
{Sewilo} M., {Watson} C., {Araya} E., {Churchwell} E., {Hofner} P., {Kurtz} S.,
  2004, \apjs, 154, 553

\bibitem[{{Shirley} {et~al}\mbox{.}(2013){Shirley}, {Ellsworth-Bowers},
  {Svoboda}, {Schlingman}, {Ginsburg}, {Rosolowsky}, {Gerner}, {Mairs},
  {Battersby}, {Stringfellow}, {Dunham}, {Glenn}, \& {Bally}}]{shi13}
{Shirley} Y.~L. {et~al.}, 2013, \apjs, 209, 2

\bibitem[{{Stil} {et~al}\mbox{.}(2006){Stil}, {Taylor}, {Dickey}, {Kavars},
  {Martin}, {Rothwell}, {Boothroyd}, {Lockman}, \& {McClure-Griffiths}}]{VGPS}
{Stil} J.~M. {et~al.}, 2006, \aj, 132, 1158

\bibitem[{{Strong} \& {Mattox}(1996)}]{sm96}
{Strong} A.~W., {Mattox} J.~R., 1996, \aap, 308, L21

\bibitem[{{Tokunaga}(2000)}]{tok00}
{Tokunaga} A.~T., 2000, {Infrared Astronomy}, {Cox} A.~N., ed., p. 143

\bibitem[{{Weaver} {et~al}\mbox{.}(1977){Weaver}, {McCray}, {Castor},
  {Shapiro}, \& {Moore}}]{wea77}
{Weaver} R., {McCray} R., {Castor} J., {Shapiro} P., {Moore} R., 1977, \apj,
  218, 377

\bibitem[{{Werner} {et~al}\mbox{.}(2004){Werner}, {Roellig}, {Low}, {Rieke},
  {Rieke}, {Hoffmann}, {Young}, {Houck}, {Brandl}, {Fazio}, {Hora}, {Gehrz},
  {Helou}, {Soifer}, {Stauffer}, {Keene}, {Eisenhardt}, {Gallagher}, {Gautier},
  {Irace}, {Lawrence}, {Simmons}, {Van Cleve}, {Jura}, {Wright}, \&
  {Cruikshank}}]{wer04}
{Werner} M.~W. {et~al.}, 2004, \apjs, 154, 1

\bibitem[{{Whittet}(2003)}]{whi03}
{Whittet} D.~C.~B., ed., 2003, {Dust in the galactic environment}

\bibitem[{{Wright} {et~al}\mbox{.}(2010){Wright}, {Eisenhardt}, {Mainzer},
  {Ressler}, {Cutri}, {Jarrett}, {Kirkpatrick}, {Padgett}, {McMillan},
  {Skrutskie}, {Stanford}, {Cohen}, {Walker}, {Mather}, {Leisawitz}, {Gautier},
  {McLean}, {Benford}, {Lonsdale}, {Blain}, {Mendez}, {Irace}, {Duval}, {Liu},
  {Royer}, {Heinrichsen}, {Howard}, {Shannon}, {Kendall}, {Walsh}, {Larsen},
  {Cardon}, {Schick}, {Schwalm}, {Abid}, {Fabinsky}, {Naes}, \& {Tsai}}]{wri10}
{Wright} E.~L. {et~al.}, 2010, \aj, 140, 1868

\bibitem[{{Zinnecker} \& {Yorke}(2007)}]{zin07}
{Zinnecker} H., {Yorke} H.~W., 2007, \araa, 45, 481

\end{thebibliography}

\IfFileExists{\jobname.bbl}{}
{\typeout{}
\typeout{****************************************************}
\typeout{****************************************************}
\typeout{** Please run "bibtex \jobname" to optain}
\typeout{** the bibliography and then re-run LaTeX}
\typeout{** twice to fix the references!}
\typeout{****************************************************}
\typeout{****************************************************}
\typeout{}
}

\end{document}